\documentclass[preprint,12pt]{elsarticle}

\usepackage{amssymb}
\usepackage{amsmath}
\usepackage{lineno}
\usepackage{xcolor}  
\usepackage{hyperref}

\usepackage{physics}
\usepackage{soul}
\usepackage{textcomp}
\usepackage{gensymb}
\usepackage{graphicx}

\journal{Astroparticle Physics}

\begin{document}
\begin{frontmatter}

\title{Reconstruction of inclined extensive air showers using radio signals: from arrival times and amplitudes to direction and energy} 

\author[lpnhe,iap]{Marion Guelfand} 
\author[subatech]{Valentin Decoene}
\author[lpnhe,iap]{Olivier Martineau-Huynh}
\author[lagrange]{Simon Prunet}
\author[conicet]{Mat\'ias Tueros}
\author[SFSU]{Oscar Macias}
\author[CEA]{Aurélien Benoit-Lévy}

\affiliation[lpnhe]{organization={ Sorbonne Université, CNRS, Laboratoire de Physique Nucléaire et des Hautes Energies (LPNHE)},
            addressline={4 Pl. Jussieu}, 
            city={Paris},
            postcode={75005}, 
            country={France}}

\affiliation[iap]{organization={ Sorbonne Université, CNRS, Institut d'Astrophysique de Paris (IAP)},
            addressline={98bis, Blvd Arago}, 
            city={Paris},
            postcode={75014}, 
            country={France}}
            
\affiliation[subatech]{organization={SUBATECH, IN2P3-CNRS, Nantes Université, Ecole des Mines de Nantes},
            addressline={4 rue Alfred Kaslter}, 
            city={Nantes},
            postcode={44300}, 
            country={France}}

\affiliation[lagrange]{organization={Université Côte d’Azur, Observatoire de la Côte d’Azur, CNRS, Laboratoire Lagrange},
            addressline={Bd de l’Observatoire}, 
            city={Nice},
            postcode={34229}, 
            country={France}}

\affiliation[conicet]{organization={IFLP - CCT La Plata - CONICET, Depto. de Física, Fac. de Cs. Ex., Universidad Nacional de La Plata},
            addressline={67 Casilla de Correo}, 
            city={La Plata},
            postcode={727 (1900)}, 
            country={Argentina}}

\affiliation[SFSU]{organization={Department of Physics and Astronomy, San Francisco State University},
            addressline={1600 Holloway Ave}, 
            city={San Francisco},
            postcode={94132}, 
            state={California},
            country={USA}}
\affiliation[CEA]
{organization={Université Paris-Saclay, CEA, List}, postcode={F-91120}, city={Palaiseau}, country={France}}

\begin{abstract}
Radio detection is now an established technique for the study of ultra-high-energy (UHE) cosmic rays with energies above $\sim10^{17}$\,eV. 
The next-generation of radio experiments aims to extend this technique to the observation of UHE earth-skimming neutrinos, which requires the detection of very inclined extensive air showers (EAS).
In this article we present a new reconstruction method for the arrival direction and the energy of EAS. It combines a point-source-like description of the radio wavefront with a phenomenological model: the Angular Distribution Function (ADF). The ADF describes the angular distribution of the radio signal amplitude in the 50-200\,MHz frequency range, with a particular focus on the Cherenkov angle, a crucial feature of the radio amplitude pattern. The method is applicable to showers with zenith angles larger than $60\degree$, and in principle up to neutrino-induced showers with up-going trajectories. It is tested here on a simulated data set of EAS induced by cosmic rays. A resolution better than 4 arc-minutes ($0.07\degree$) is achieved on arrival direction, as well as an intrinsic resolution of 5\% on the electromagnetic energy, and around 15\% on the primary energy.
\end{abstract}


\end{frontmatter}

\tableofcontents
\section{Introduction}
Several experiments (AERA \cite{Abreu_2012}, CODALEMA \cite{ARDOUIN_2005}, LOPES~\cite{Huege_2008}, LOFAR \cite{LOFAR}, Tunka-Rex\cite{Tunka-Rex:2019msu}) have demonstrated that radio detection is a robust and efficient technique to reconstruct properties of ultra-high-energy cosmic rays (UHECRs) for zenith angles up to $60\degree$. These experiments exploit the fact that the extensive air shower (EAS) induced by an ultra-high energy (UHE) cosmic particle when it enters the atmosphere is associated with electromagnetic radiation. This transient emission is caused mostly by the drift of electrons and positrons from the EAS under the influence of the Earth's magnetic field. Given the $\sim$meter thickness of the air shower, it is coherent in the 10-100\,MHz range at emission, and up to the GHz range thanks to propagation effects. For primary particles with energies larger than a few $10^{16}$\,eV, the electromagnetic pulse  can reach amplitudes larger than the dominant sources of noise at these frequencies, in particular from the Galaxy, and the ground blackbody radiation.

Among present or future projects for EAS radio detection, several target very inclined air showers using extended arrays (e.g. AugerPrime radio upgrade \cite{Castellina_2019}, GRAND \cite{Alvarez_Muniz_2019},  GCOS\cite{Coleman_2023}).
These arrays capitalize on the fact that footprints of inclined air showers span several tens of square kilometers, enabling good detection efficiency with low antenna density. This allows for the deployment of sparse arrays over large surfaces and thus enables the detection of the lowest astroparticle flux expected at the highest energies.

The Giant Radio Array for Neutrino Detection (GRAND) is an envisioned multi-messenger observatory that primarily targets Earth-skimming air showers induced by UHE neutrinos \cite{Alvarez_Muniz_2019}. Such a goal requires dedicated reconstruction methods to derive relevant information --direction of origin, energy and nature-- of cosmic primaries from the radio data. Existing reconstruction methods were initially developed for cosmic-ray-induced air showers with zenith angles below 60$\degree$ and become much less efficient for very inclined air showers. The latter correspond to geometries and environments for shower developments which differ significantly from the vertical case (defined in this article as zenith angle $\in [0\degree- 60\degree]$): in particular thinner air at shower maximum in the case of cosmic-ray-induced showers with large zenith, and much thicker in the case of neutrinos, affect the characteristics of radio emission and propagation. Some energy reconstruction methods were later extended down to $\sim85\degree$ \cite{Schlüter_2023} but rely on the shower core position. In their present implementation, such methods can not be applied for upward-going showers induced by neutrino interactions with the Earth target, where no point of impact between the air shower axis and the ground exists. Finally, current methods for determining the arrival direction lack the precision needed to achieve a point-degree angular resolution. 

In this article, we propose a method to reconstruct the EAS direction of origin and energy which does not rely on the shower core position. 
It is tested here on downward-going cosmic-ray-induced air shower and can in principle be extended to up-going showers associated with neutrinos.
The method combines plane and spherical reconstructions of the shower wavefront with what we refer to as the \textit{Angular Distribution Function} (ADF). This analytical function is tailored to match the amplitude distribution of radio signals in a two-dimensional angular space. Using this approach, we can accurately reconstruct the direction of the shower's origin. Additionally, because the amplitude of the radio signal is directly related to the electromagnetic energy of the air shower, this method also allows us to infer the energy of the air shower itself.

The paper is structured as follows: in Section~\ref{section:footprint_pattern}, we detail the different terms of the ADF model. We discuss in Section~\ref{section:cherenkov_angle} the range of validity of the model, in particular by studying the precision of the ADF prediction of the amplitude profile maxima. We present in Section~\ref{section:reconstruction} the complete pipeline for arrival direction reconstruction and apply the method to a set of realistic simulations. Finally, in Section \ref{section:energy}, we detail how the electromagnetic energy can be reconstructed, and evaluate the resolution of the method on this parameter and the primary energy reconstruction with a set of realistic simulations. 

\section{Study of the air-shower amplitude distribution}
\label{section:footprint_pattern}
We first give a brief summary of the principles of electromagnetic emission by air showers. A more complete review can for instance be found in\,\cite{Huege:2016veh, Schroder:2016hrv}. EAS electromagnetic emission in the MHz frequency range results from the combination of two dominant emission mechanisms\,\cite{KahnLerch_1966}: i) the drift of the charged particles from the EAS in the Earth magnetic field, called geomagnetic emission and linearly polarized in the Lorentz force direction; ii) the moving dipole, from the accumulation of negative charged particles in the front of the shower, called Askaryan (or charge-excess) emission and radially polarized from the shower axis. The interplay of these two emission mechanisms leads to a polarization pattern called the geomagnetic asymmetry, where the contributions of the two emission mechanisms have coincident or opposite polarizations, and are thus added or subtracted along the Lorentz force direction. In addition, the propagation of electromagnetic waves in the atmosphere, where the refractive index is larger than $1$, leads to compression of the radio signal in the time domain along a cone centered on the shower direction which we call the radio Cherenkov compression. Finally, for inclined EAS the relative distance to the emission region differs significantly among antennas: "early" antennas closer to the shower emission region measure larger signal amplitudes than "late" antennas\,\cite{Schlüter_2023}.

\subsection{Angular Distribution Function (ADF) coordinate system}
\label{section:coordinates}
\begin{figure}
    \centering
    \includegraphics[width=0.8\linewidth]{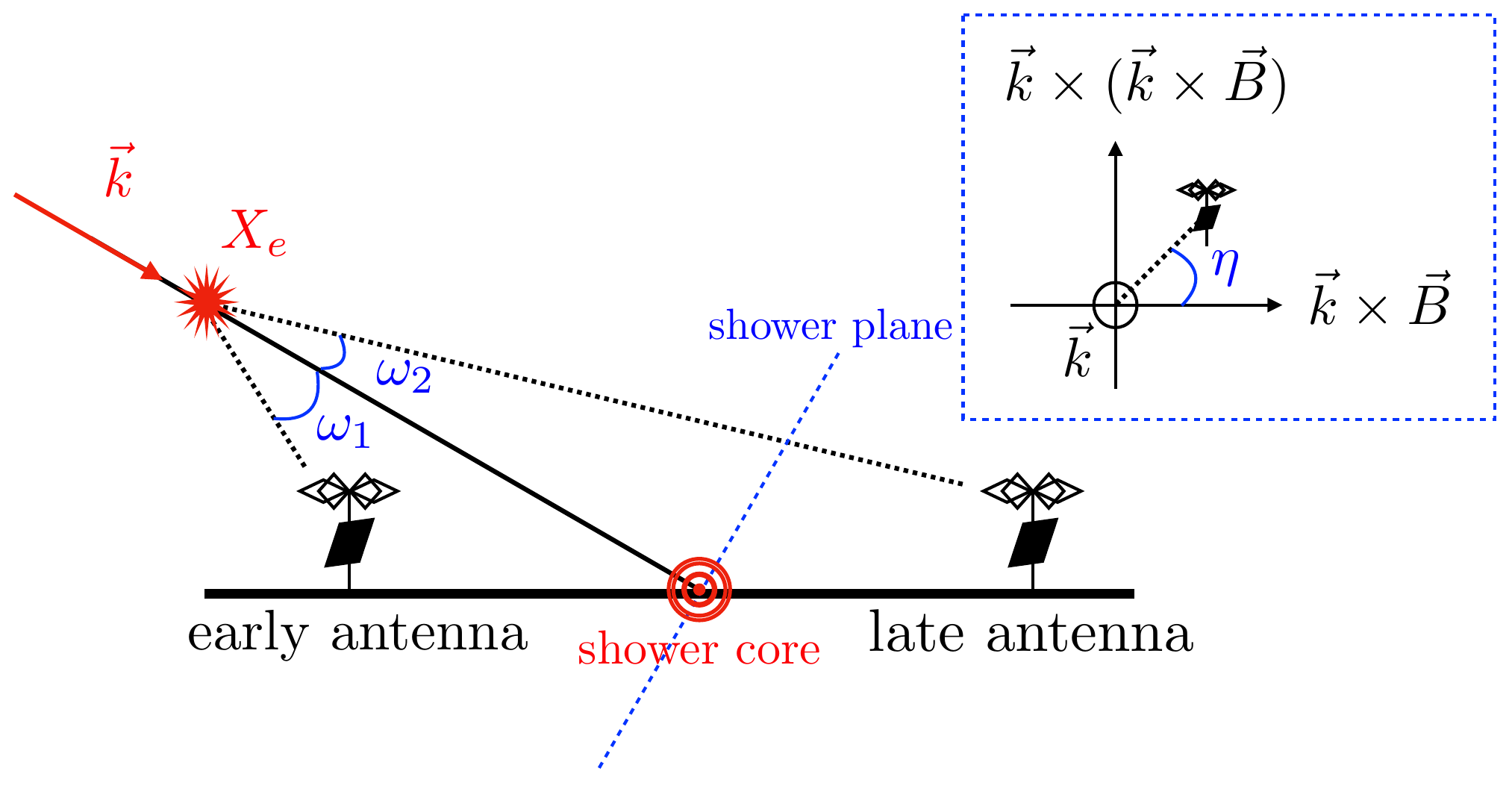}
   \caption{ Sketch of the ADF coordinate system. The $\omega_{i}$ angle is the angular distance between each antenna and the shower axis, measured from the $X_{e}$ emission point (see text for details). The $\eta$  angle is the angular position of the antenna with respect to the vector $\mathbf{k} \times \mathbf{B}$ and goes from $0\degree$, when aligned with the positive $\mathbf{k} \times \mathbf{B}$ direction, to $180\degree$, when aligned with the negative $\mathbf{k} \times \mathbf{B}$ direction.}
   
    \label{fig:sketch_adf_coordinates_systems}
\end{figure}

We describe the EAS amplitude distribution through the angular coordinates ($\omega, \eta$) of the antennas, where $\omega$ is the angular distance to the shower axis measured from a reference position $X_{\rm e}$ and $\eta$ the rotation angle measured from the $(\mathbf{k} \times \mathbf{B})$ axis in the so-called shower plane, with $\mathbf{k}$ the shower direction unit vector and $\mathbf{B}$ the unit vector in the direction of the geomagnetic field. This plane is perpendicular to the shower axis, and defined by the basis \{$\mathbf{k} \times \mathbf{B}, \mathbf{k} \times (\mathbf{k} \times \mathbf{B})$\} (see Figure~\ref{fig:sketch_adf_coordinates_systems}).

The reference $X_{\rm e}$ is the source point of a spherical wavefront adjusted to the signal arrival times at the antenna locations. 
It was shown in \cite{decoene_2023} that for a sparse array (with a $\sim$kilometer step) and a time resolution of the order of the GPS jitter (typically $\gtrsim 5$\,ns), a spherical approximation of the EAS wavefront is valid for zenith larger than $60\degree$. In this case, the signal time information does not allow to distinguish the EAS radio emission from a point-like model.
The same (strong) assumption is made for the ADF treatment presented here, and the point-like location of the EAS radio emission is used to define the angle $\omega$ in the following. The validity of this assumption within the context of the ADF model will be evaluated in the next section. 

The motivation to use angular coordinates is that --despite second-order effects that will be later studied-- EAS exhibits a conical geometry, which is better treated with angular coordinates rather than linear ones. 
Moreover, this choice of coordinates scales more naturally with different shower inclinations since the projection on the ground of the radio emission is straightforward with the $\omega$ angles. Finally, the choice to set the origin of our frame at an emission point instead of the shower core (as it is traditionally done) allows us to equally describe downward-going EAS and upward-going EAS induced by neutrinos.

\subsection{Description of the ADF}
\label{section:description}
The Angular Distribution Function describes the maximum amplitude of the electric field produced by the EAS at any antenna position, as defined in Section \ref{section:simus}. 
It can be written as:
\begin{eqnarray}
\label{eq:adf}
    f^{\rm ADF}(\omega, \eta, \alpha, l; A, \delta \omega) = \frac{A}{l} f^{\rm Geom} (\alpha, \eta, G_{A}) f^{\rm Cherenkov} (\omega; \delta \omega),
\end{eqnarray}
here $A$ is a free parameter adjusting the amplitude, $l=l\qty(X_e, \mathbf{x}_{\rm ant})$ the longitudinal propagation distance between the $X_{\rm e}$ emission point (an input parameter here) and the antenna ($\mathbf{x}_{\rm ant}$),
and $f^{\rm Geom}$ the geomagnetic asymmetry given by:
\begin{eqnarray}
\label{eq:geomag}
    f^{\rm Geom}(\alpha, \eta, G_{A}) = 1+ G_{A} \frac{\rm cos(\eta)}{\rm sin(\alpha)}.
\end{eqnarray}
where $G_{A}$ is the geomagnetic asymmetry strength 
parameterized as a function of the zenith angle value, and $\alpha$ is the so-called geomagnetic angle (i.e. the angle between the shower propagation direction $\mathbf{k}$  and the Earth magnetic field $\mathbf{B}$).

The term $f^{\rm Cherenkov}$  is a Lorentzian parametrization of the signal enhancement observed in the radio signal around the Cherenkov angle: 
\begin{eqnarray}
\label{eq:cher}
    f^{\rm Cherenkov}(\omega, \delta \omega) = \frac{1}{1+4 \left[ \frac{(\rm tan(\omega)/ \rm tan(\omega_{c}))^{2} -1}{\delta \omega}  \right]^{2}},
\end{eqnarray}
where $\omega_{c}$ is the Cherenkov angle computed from the model presented in~\ref{section:validation} and $\delta \omega$ a free parameter of the ADF describing the width of the Cherenkov cone.

All angular variables used in the ADF model can be written explicitly as a function of the shower direction $\mathbf{k}$:
\begin{eqnarray}
\label{eq:pos}
    \omega_{i} = \rm acos(\mathbf{k},\mathbf{x}_{i}),\quad \textit{l}_{i} = \mathbf{k}\cdot \mathbf{x}_{i}, \quad \eta_{i} = \rm atan(y_{i}^{\rm sp}/x_{i}^{\rm sp}), \quad \alpha = \rm acos(\mathbf{k},\mathbf{B}),
\end{eqnarray}
where $\mathbf{x}_{i} = (x_{\rm i}, y_{\rm i}, z_{\rm i})$ is the $\rm i^{th}$ antenna position with respect to the emission source and $x_{i}^{\rm sp}, y_{i}^{\rm sp}$ its cartesian coordinates in the shower plane referential \{$\mathbf{k} \times \mathbf{B}$, $\mathbf{k} \times (\mathbf{k} \times \mathbf{B})$\}.

\subsection{Comparison to simulations}
\label{section:validation}
In this section, the various terms composing the ADF are compared against the amplitude distribution obtained from simulated data, which will first be presented. 

\subsubsection{Simulations and treatment}
\label{section:simus}
To study the ADF, we simulate air showers with Aires version 19.04.08 \cite{Sciutto_2019} and calculate their radio emission with ZHAireS \cite{Alvarez_Muniz_2012_AIRES} version 1.0.30a, using the extended Linsley's atmospheric model and an exponential model for the index of refraction, with 8.2 km scale height and 1.000325 refraction index at sea level. Sibyll 2.3d~\cite{Riehn_2019} is chosen as the hadronic model and the relative particle thinning is $10^{-5}$.

The ground altitude is set at $1086\,\rm m$ above sea level and the geomagnetic field is taken with $60.79\degree$ inclination,  $0.36\degree$ declination and total strength $55.997\,  \rm \mu T$, values corresponding to the site of the GRANDProto300 (GP300) experiment~\cite{Martineau-Huynh_hdr2021}.

\begin{figure}[!tb]
     \centering
     \includegraphics[width=0.45\columnwidth]{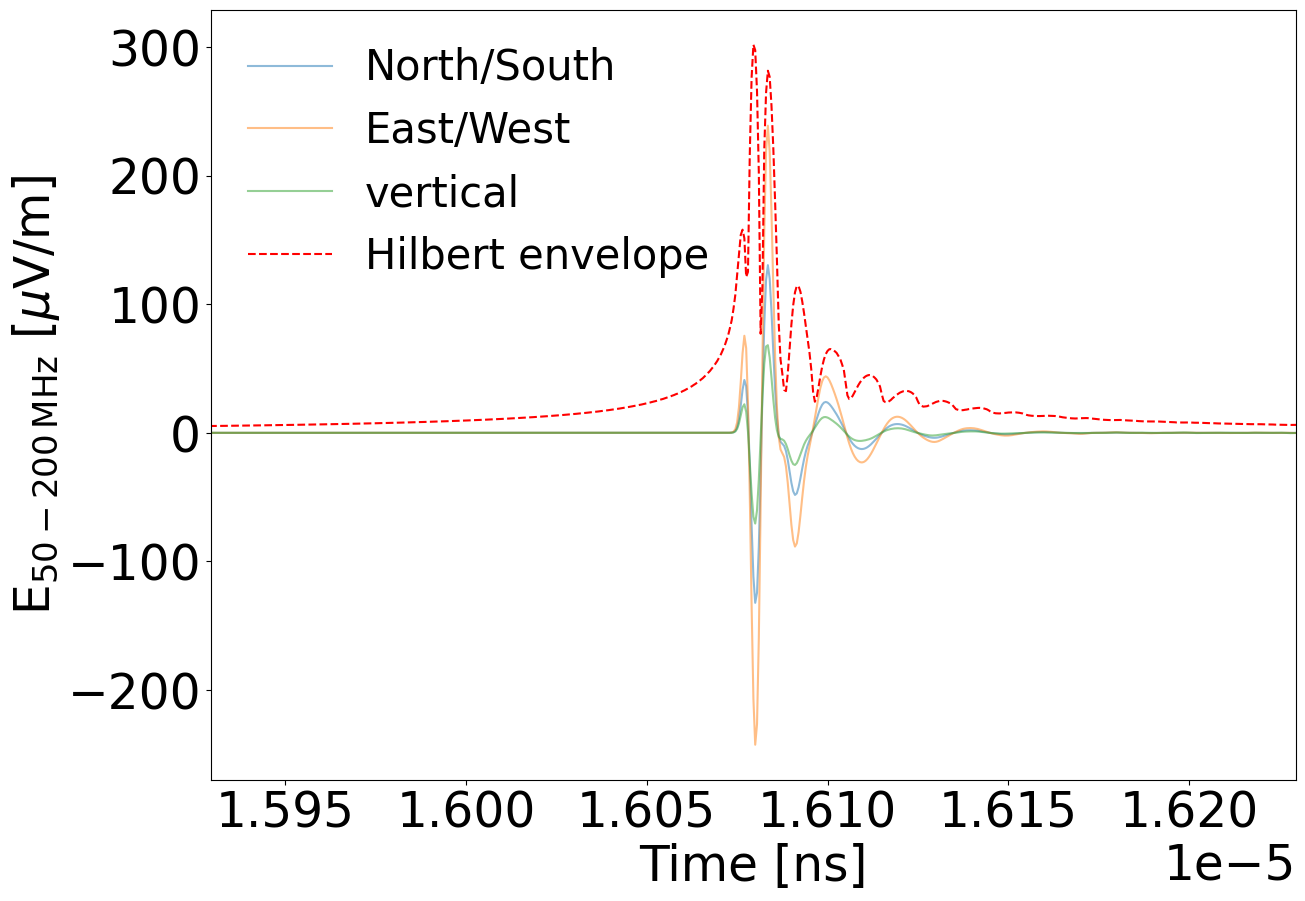}
     \includegraphics[width=0.45\columnwidth]{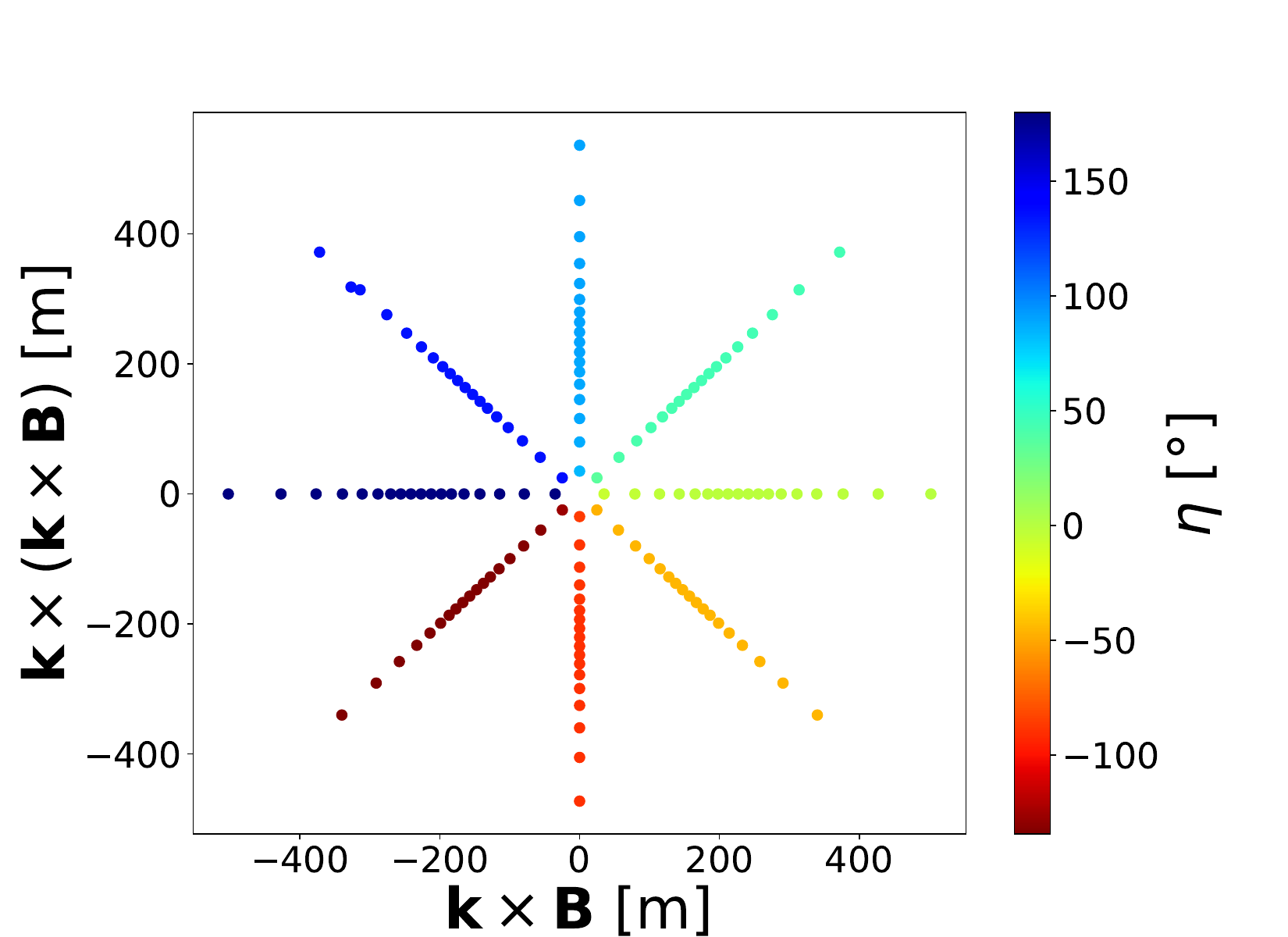}
      \caption{\textit{(Left)} Electric field trace from a ZHAireS simulation of a proton shower, filtered in the 50-200\,MHz frequency. The solid lines represent the three Cartesian components of the E-field vector and the red dotted line the Hilbert envelope computed from the norm of the electric field. \textit{(Right)} Starshape layout. The antennas are projected in the shower plane defined in the basis \{$\mathbf{k} \times \mathbf{B}, \mathbf{k} \times (\mathbf{k} \times \mathbf{B})$\}.}
    \label{fig:trace}
\end{figure}

The simulations are performed on a star-shape layout centered on the core of each event, represented in the right panel of Figure~\ref{fig:trace}, where the antennas are placed at 20 angular positions with $\omega \leq 3\degree$ along 8 arms. This array configuration is particularly suitable to detail each asymmetry effect on the data with the corresponding model component.

The data set is composed of proton showers with zenith angles distributed over logarithmic bins of $1/\rm cos(\theta)$ between $57\degree$ and $87.1\degree$ to sample homogeneously inclined to nearly horizontal showers. We used $5$ distinct azimuth angles $\phi$ ($0\degree$, $45\degree$, $90\degree$, $135\degree$, $180\degree$) to sample the geomagnetic asymmetry. The energy range is set with 22 logarithmic bins, from $0.02\,\rm EeV$, which corresponds to roughly the lower threshold for radio-detection, to $3.98\,\rm EeV$, which is a reasonable upper bound for a realistic detection rate, given the cosmic-ray fluxes and the considered detector size. 

The time series corresponding to the raw electric fields computed with ZHAireS along the three perpendicular axis (East-West, North-South and vertical) are first filtered with a Butterworth filter in the 50-200\,MHz frequency range, the bandwidth of the GRAND experiment~\cite{Alvarez_Muniz_2019}. The filtered signals are summed quadratically to compute the norm of the total electric field.  The signal peak amplitude and position in time are then computed for each antenna at the maximum of the Hilbert envelope of this signal, as shown in Figure~\ref{fig:trace}. The corresponding times are used to determine the position $X_{\rm e}$ of the radio emission point through a spherical fit of the radio wavefront. 

The peak amplitudes of the simulated data can finally be represented in the shower plane, and the ADF scaling factor $A$ and Cherenkov width $\delta \omega$ adjusted to them. These variables are used to evaluate the various terms of Eq. \ref{eq:adf}. 
In this section and the next, which aim at ADF validation, the true shower direction $\mathbf{k}$ is used in the analytical model.

\subsubsection{The Cherenkov enhancement}
\label{section:pattern}
The radio signal amplitude exhibits a specific enhancement around the Cherenkov angle (see e.g. Figure~\ref{fig:geomag}). As will be discussed in section \ref{sec:frequency_effect}, the shape of the amplitude profile evolves with EAS zenith angle and the selected radio frequency range. For the 50-200\,MHz range chosen for this study and air showers with zenith angles above $60\degree$, it was found that a Lorentzian distribution is very well suited.

The maxima positions of the term $f^{Cherenkov}$ in Equation \ref{eq:cher} are reached for angles $\omega = \pm \omega_c$. This could be left as a free parameter of the ADF fit for simplicity, yet it should be noted that it is directly linked to the physical process of EAS radio emission and propagation. It is therefore 
possible to compute $\omega_c$, as will be discussed in section \ref{section:cherenkov_angle}.

\subsubsection{The geomagnetic asymmetry}
\label{section:geomag}
\begin{figure}[!ht]
     \centering
     \includegraphics[width=0.49\columnwidth]{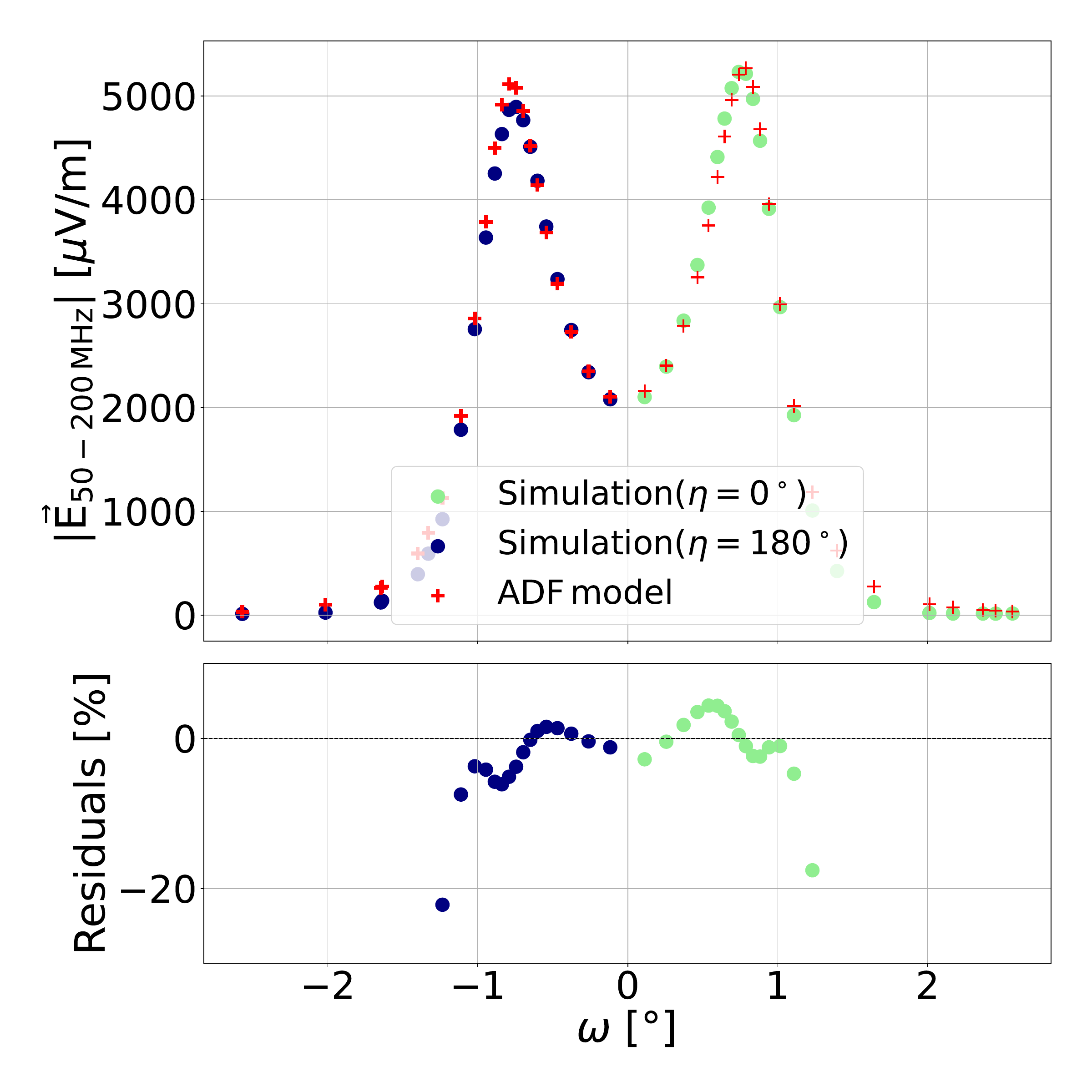}
      \includegraphics[width=0.49\columnwidth]{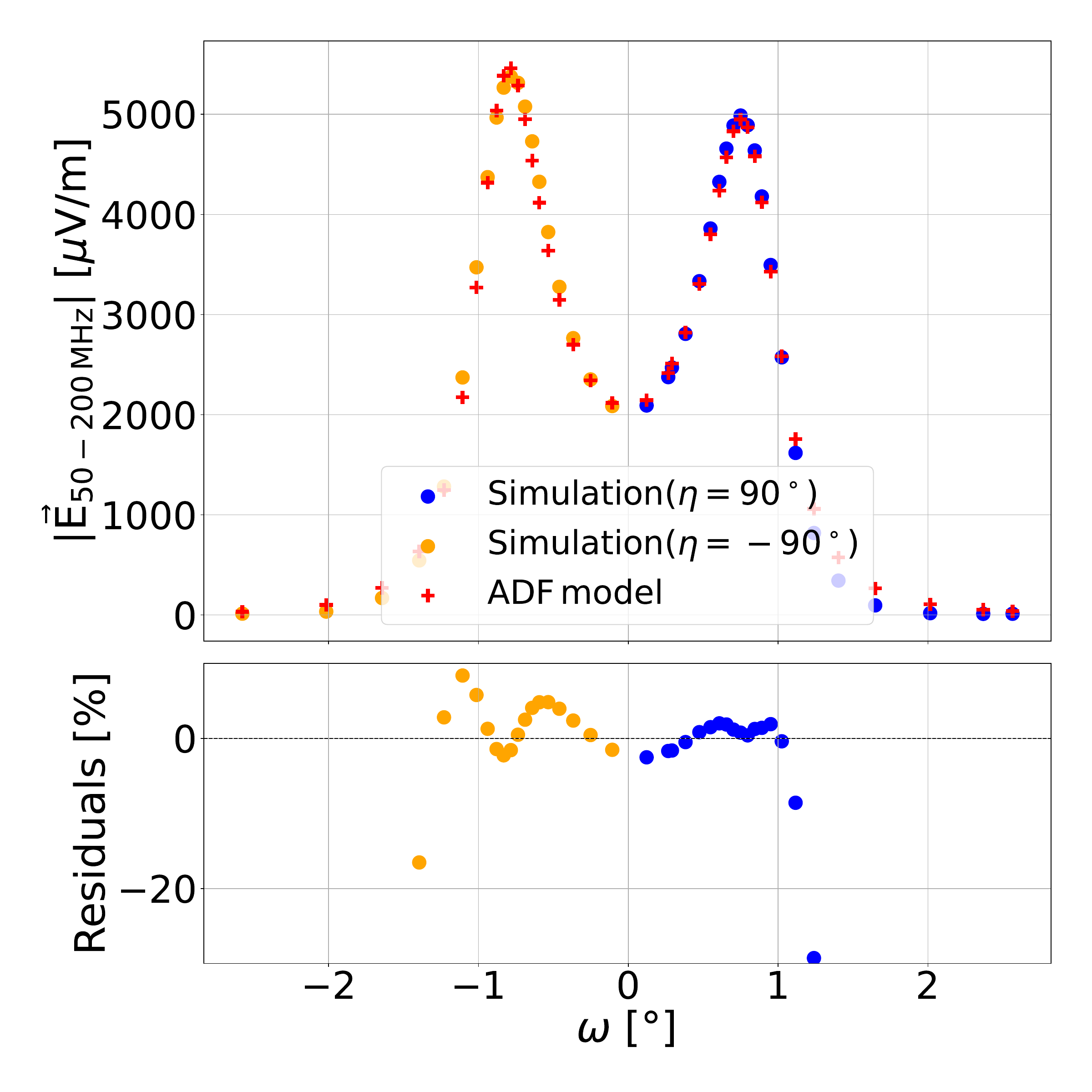}
      \caption{(\textit{Left}) Illustration of the geomagnetic effect. The signal amplitude distribution is displayed along the $\mathbf{k} \times \mathbf{B}$ axis. Antennas with $\eta = 0\degree$ (green) receive more signal than antennas with $\eta = 180\degree$ (dark blue). (\textit{Right}) Illustration of the early-late effect. The signal amplitude distribution is displayed along the $\mathbf{k} \times (\mathbf{k} \times \mathbf{B})$ axis. Early antennas (orange, $\eta = -90\degree$) are closer to the emission zone of the shower, than late  antennas (blue, $\eta = 90\degree$) and therefore exhibit a larger amplitude. Signal computed for a EAS with $\theta = 74.76\degree$, $\phi = 180\degree$, primary proton of energy 3.98 EeV. For both plots the adjusted ADF is displayed with red crosses. The bottom panels display the relative residuals, defined as the relative difference between the simulated signal amplitude and the modeled signal amplitude at each $\omega$ position.}
    \label{fig:geomag}
\end{figure}

The radio footprint exhibits a difference in amplitude along the $\mathbf{k} \times \mathbf{B}$ axis, with larger values for antennas in the range $\eta \in$ [-90°,+90°] (see Figure~\ref{fig:geomag}, left). This originates from the interplay between charge-excess and geomagnetic emission mechanisms, associated with two different polarization patterns~\cite{KahnLerch_1966}. The geomagnetic effect is highly dominant for strong magnetic fields such as the one chosen for this study. The same effect is also expected for inclined trajectories, when showers develop in thinner atmospheres~\cite{Chiche_2022}. Geomagnetic asymmetry is, therefore, a minor correction and is taken into account here for the sake of completeness only. 

We determine the term $G_A$ from  Eq.~\ref{eq:geomag} by computing and normalizing the difference measured for our simulated EAS between the two peak amplitudes at $\eta = 0\degree$ and  $\eta = 180\degree$ (for which the geomagnetic asymmetry is maximal). In the zenith range [$57\degree, 87\degree$], this geomagnetic asymmetry could be fit by a linear law\,:
\begin{eqnarray}
\label{eq:geom_param}
G_A = -0.0026\times\theta+0.220
\end{eqnarray}
where $\theta$ is the zenith angle in degrees. This parametrization aligns with the results presented in~\cite{Chiche_2022}, though derived differently.

\subsubsection{The early-late effect}
\label{section:early-late}
The electric field amplitude of the radio emission decreases during propagation through the atmosphere. From basic principles of energy conservation, this effect can be described as a $\propto 1/l$ pre-factor in Eq.~\ref{eq:adf}. For non-vertical air showers, 
as illustrated on Figure~\ref{fig:sketch_adf_coordinates_systems}, the "early" antennas are closer to the $X_{e}$ emission point, and will measure a stronger signal than "late" antennas located further away to $X_{e}$.
This asymmetry, called the early-late effect \cite{Schlüter_2023}, is taken into account in our model with the $A/l$ term in Eq.~\ref{eq:adf}, as shown in the right panel of Figure~\ref{fig:geomag}. \\

We explained the various terms of the ADF and motivated them with simulated data.
In the next section, we will focus specifically on the angle $\omega_c$ from Eq. \ref{eq:cher}, a key parameter of the ADF model and cornerstone of our method. We explain how it is computed and how it relates to the angular position of the amplitude maximum for our simulated dataset.

\section{Study of the ADF maxima}
\label{section:cherenkov_angle}
We use for this study the simulation set described in section \ref{section:simus}, restricting it to the 566 air showers with azimuth angles $\phi = 0\degree$ (originating from North) and $\phi = 180\degree$ (originating from South), as for these specific azimuth angles the Cherenkov asymmetry is expected to be maximal along the $\mathbf{k} \times (\mathbf{k} \times \mathbf{B})$ axis and null along the $\mathbf{k} \times \mathbf{B}$ axis, allowing for an easier study of this effect.

As in the previous section, the true value of $\mathbf{k}$ is used, and only the scaling factor $A$ and Cherenkov width $\delta \omega$ of the ADF are adjusted to the simulated data. 

\subsection{Maxima of the air-shower radio amplitude profile}

The radio signal amplitude induced by EAS exhibits a specific enhancement around the Cherenkov angle~\cite{Nelles_2014}. This experimental feature is well reproduced with simulations~\cite{AlvarezMuniz_2010} and is understood as a consequence of the compression of the signal in the time domain during propagation in a medium with refractive index $\neq$ 1~\cite{deVries_2011}. The exact position, shape, and amplitude of this feature depend on frequency, as discussed in section \ref{sec:frequency_effect}, but at first order, the effect is expected to be maximal along the Cherenkov angle. At emission level, and in a homogeneous medium with a constant refractive index $n$, emission coherence gives a Cherenkov angle value:
\begin{eqnarray}
\label{eq:cherenkov_standard}
\omega_{c} = {\rm acos}(1/n).    
\end{eqnarray}
Yet observers do not stand at emission level and their respective optical paths depend on their actual position. In addition, the refractive index varies in the atmosphere. In the ZHAireS simulation program, this is modelled using an exponential profile, which decreases with increasing altitude, referred to as $n\qty(h)$. This refractive index is used to compute the electromagnetic emission induced by air showers at different altitudes and is computed assuming a spherical Earth curvature model. The propagation between the emission and the observer is handled assuming optical paths following straight lines, i.e. it does not take into account light bending caused by the variable refractive index. However, propagation time delays are computed using an effective refractive index $n_{\rm eff}\qty(R)$ defined as the average value of the index of refraction along the path (see~\cite{Alvarez_Muniz_2012_AIRES} for more details).
This guarantees that the propagation times and wavefront shapes are realistic.

\begin{figure}[!tb]
     \centering
     \includegraphics[height=0.5\columnwidth]{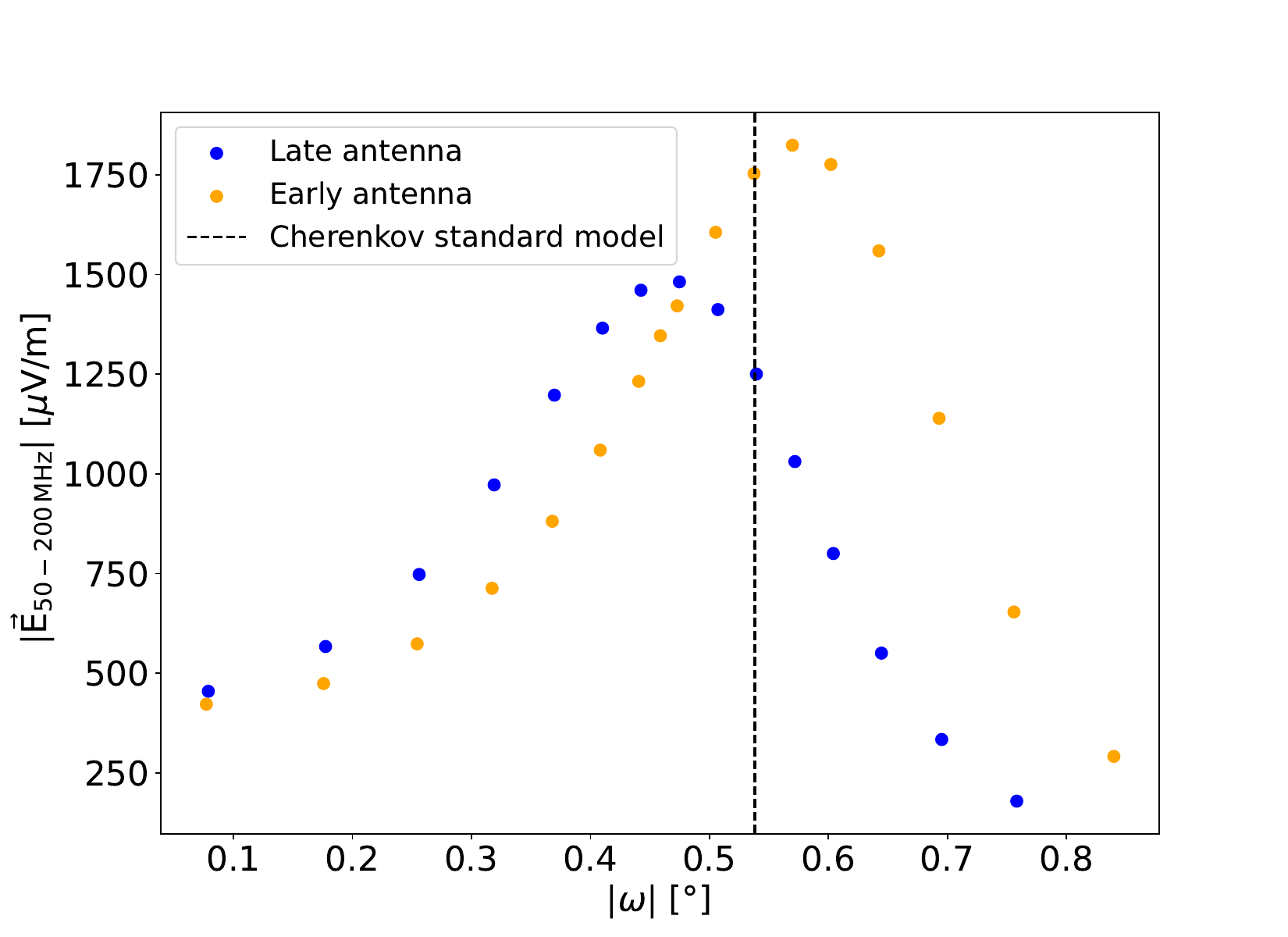}
      \caption{Illustration of the Cherenkov asymmetry. The signal amplitude distribution is displayed along the $\mathbf{k} \times (\mathbf{k} \times \mathbf{B})$ axis for primary proton of energy $3.16\, \rm EeV$, $\phi = 180\degree$ and $\theta = 84.95\degree$. The amplitude maximum is further away from the shower axis for early antennas. The dark dotted line represents the expected Cherenkov angle value for the refractive index at the $X_{e}$ altitude.}
    \label{fig:cherenkov_assym}
\end{figure}

It is observed in ZHAireS simulations that, for inclined air showers, the angular position of the amplitude peak is slightly larger for early antennas than for late ones, as shown in Figure~\ref{fig:cherenkov_assym} and already documented in \cite{Schluter_2020}. The Cherenkov angle computed from Eq.\,\ref{eq:cherenkov_standard} with the refractive index taken at the emission position $X_{e}$  neither reproduces the observed asymmetry nor corresponds to the position of any of the two peaks. This is also verified when considering the emission at $X_{\rm max}$, which shows the limitation of this standard computation.

\label{sec:inclination_effect}

\subsection{The Cherenkov asymmetry toy model}
To model this effect, we have developed a toy computation based on the two following assumptions (i) an observer placed along the Cherenkov angle experiences the maximal time compression (i.e., the time delay between the instants of arrival of emissions from different parts of the shower is minimal), hence resulting in a strongly peaked signal in the time domain. (ii) the EAS radiation is emitted within a few kilometers~\cite{guelfand_2024} around the emission point $X_e$. 

\begin{figure}
    \centering
    \includegraphics[width=0.8\linewidth]{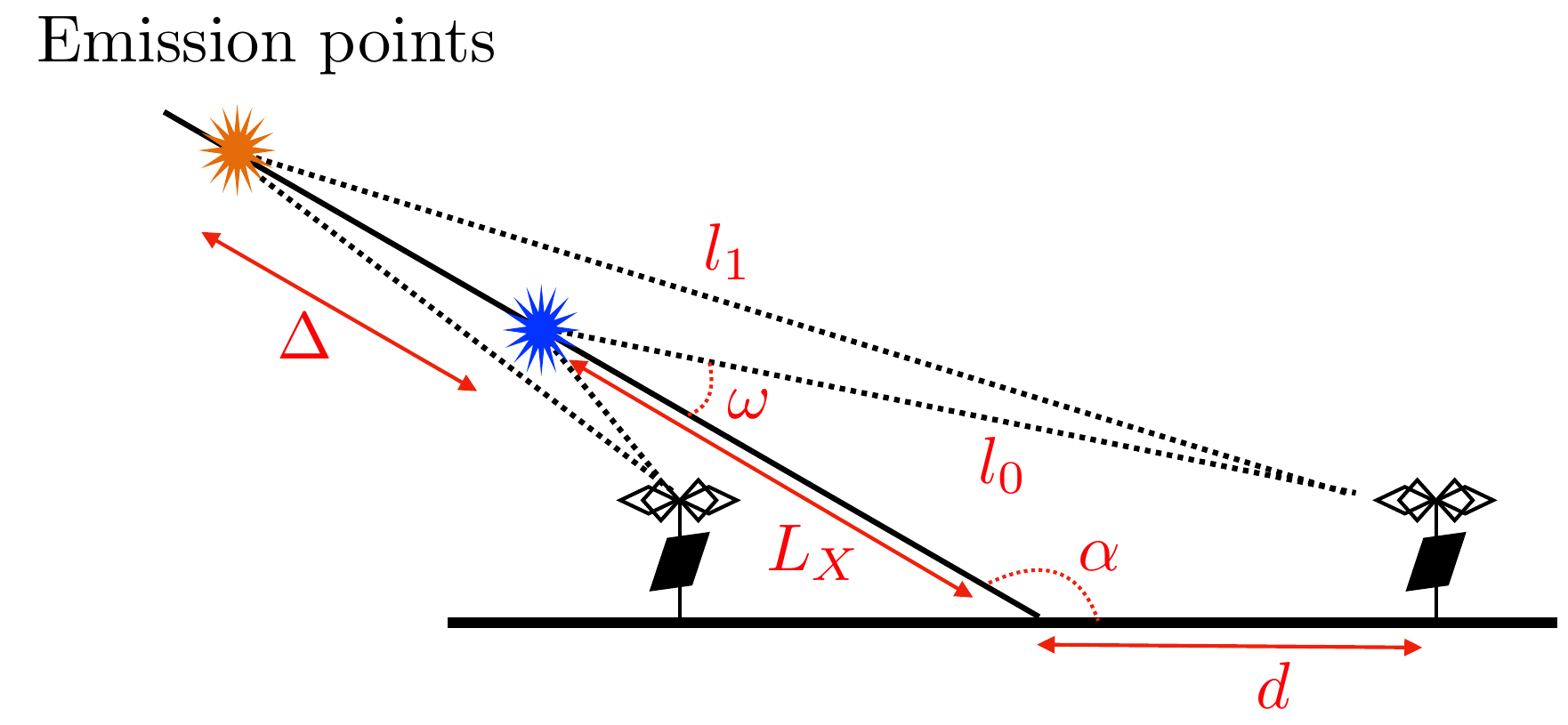}
    \caption{ Sketch of the Cherenkov asymmetry model, which describes shower emission with two points separated by a few kilometers along the shower axis. The Cherenkov angle then corresponds to ground positions for which optical paths difference from these two points is minimal.}
    \label{fig:sketch_cherenkov_model}
\end{figure}

\begin{figure}[!ht]
         \includegraphics[width=0.5\columnwidth]{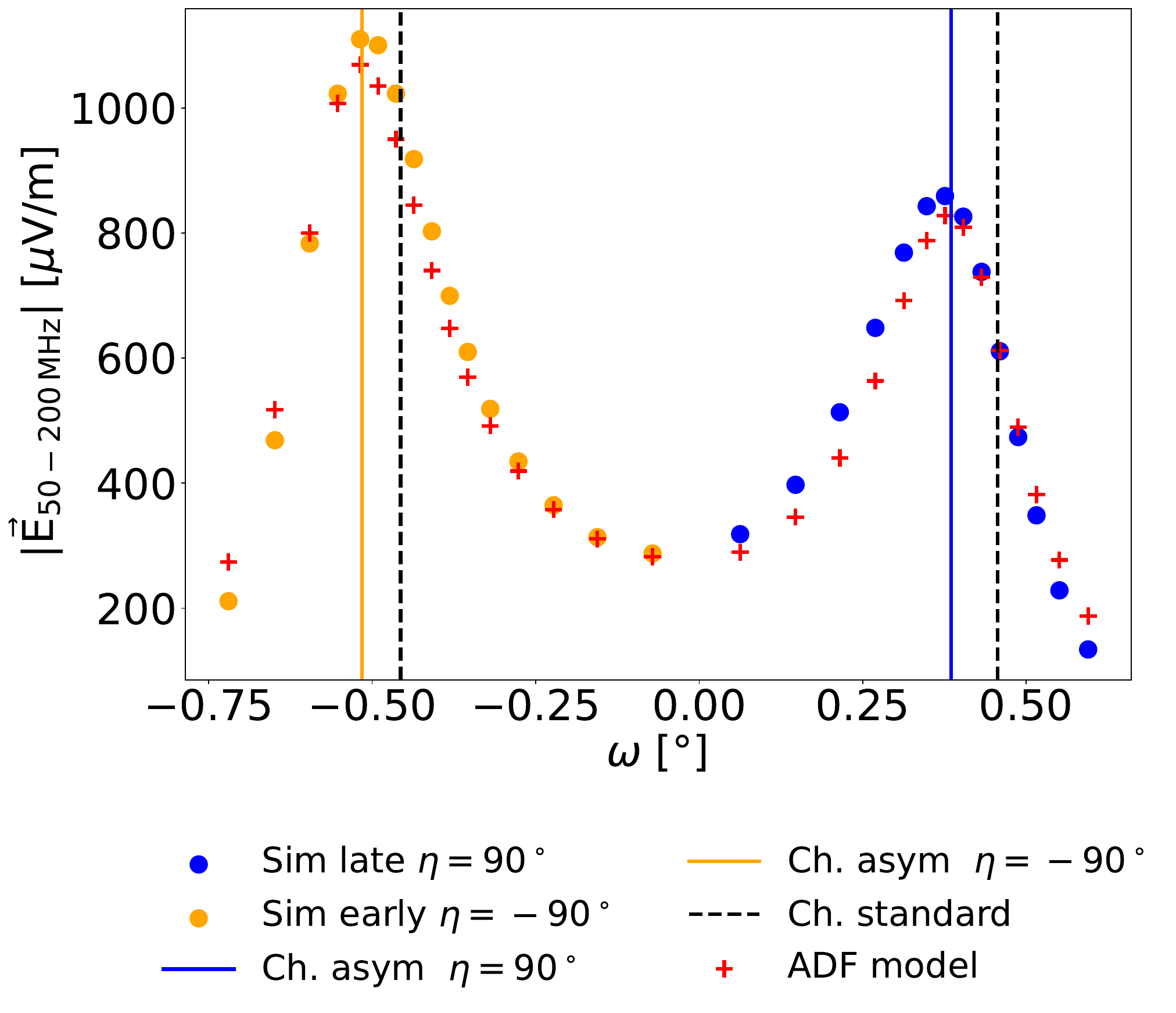}
         \includegraphics[width=0.5\columnwidth]{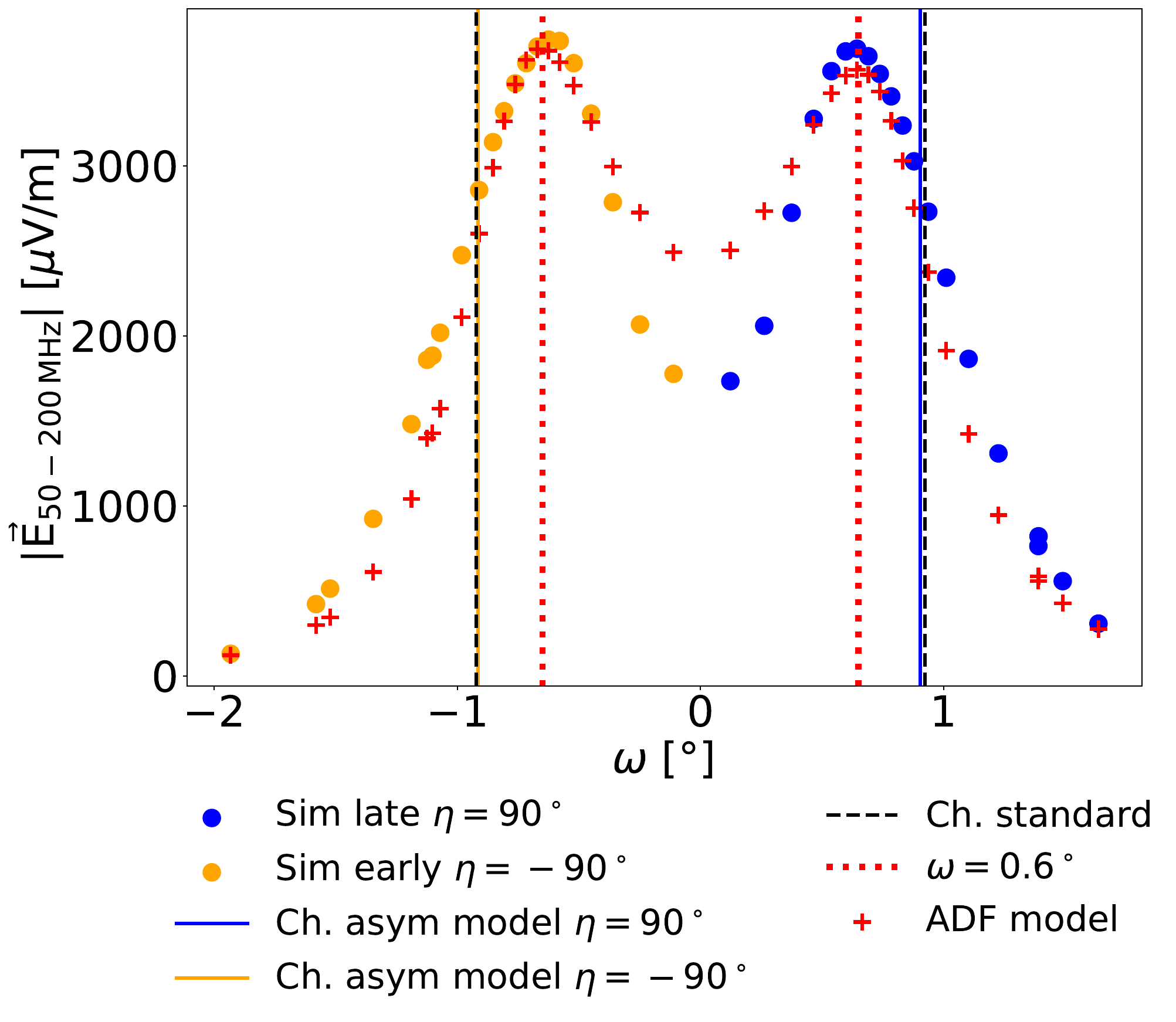}
     \caption{Signal amplitude distribution along the $\mathbf{k} \times (\mathbf{k} \times \mathbf{B})$ axis for primary proton of energy $3.16\, \rm EeV$, $\phi = 0\degree$ and (\textit{Left}) $\theta = 86.5\degree$ and (\textit{Right}) $\theta = 56.95\degree$. Above $70\degree$, the Cherenkov angle computed with the asymmetry toy model for late (early) antennas is represented with a blue (orange) dotted line. Below $70\degree$, the Cherenkov angle is computed as the minimum between the value obtained with the asymmetry toy model and $0.6\degree$ (as in this example). The resulting value is then included in the ADF model as the parameter $\omega_c$, and the resulting ADF is plotted as red crosses. The standard Cherenkov angle is also represented with a black dotted line and we clearly see that it fails to reproduce the Cherenkov angle from simulated data.}
      \label{fig:cherenkov_assym_model}
\end{figure}

Our toy model hence simply consists in computing semi-analytically the angle which minimizes the light path difference between two points symmetrically placed  around $X_e$, as sketched in Figure~\ref{fig:sketch_cherenkov_model}, using here the same refractive index computation as in ZHAireS simulations. The computation is detailed in~\ref{sec:annex:TMCherenkov}, and the resulting Cherenkov angle is shown in the left panel of Figure~\ref{fig:cherenkov_assym_model}
for an inclined shower ($\theta = 86.5\degree$). The right panel, which displays a vertical air shower ($\theta = 56.95\degree$) will be discussed in Section~\ref{sec:systematic_effects}, after presenting the systematic errors we encountered and how we addressed them.

\begin{figure}[!ht]
         \includegraphics[width=0.5\columnwidth]{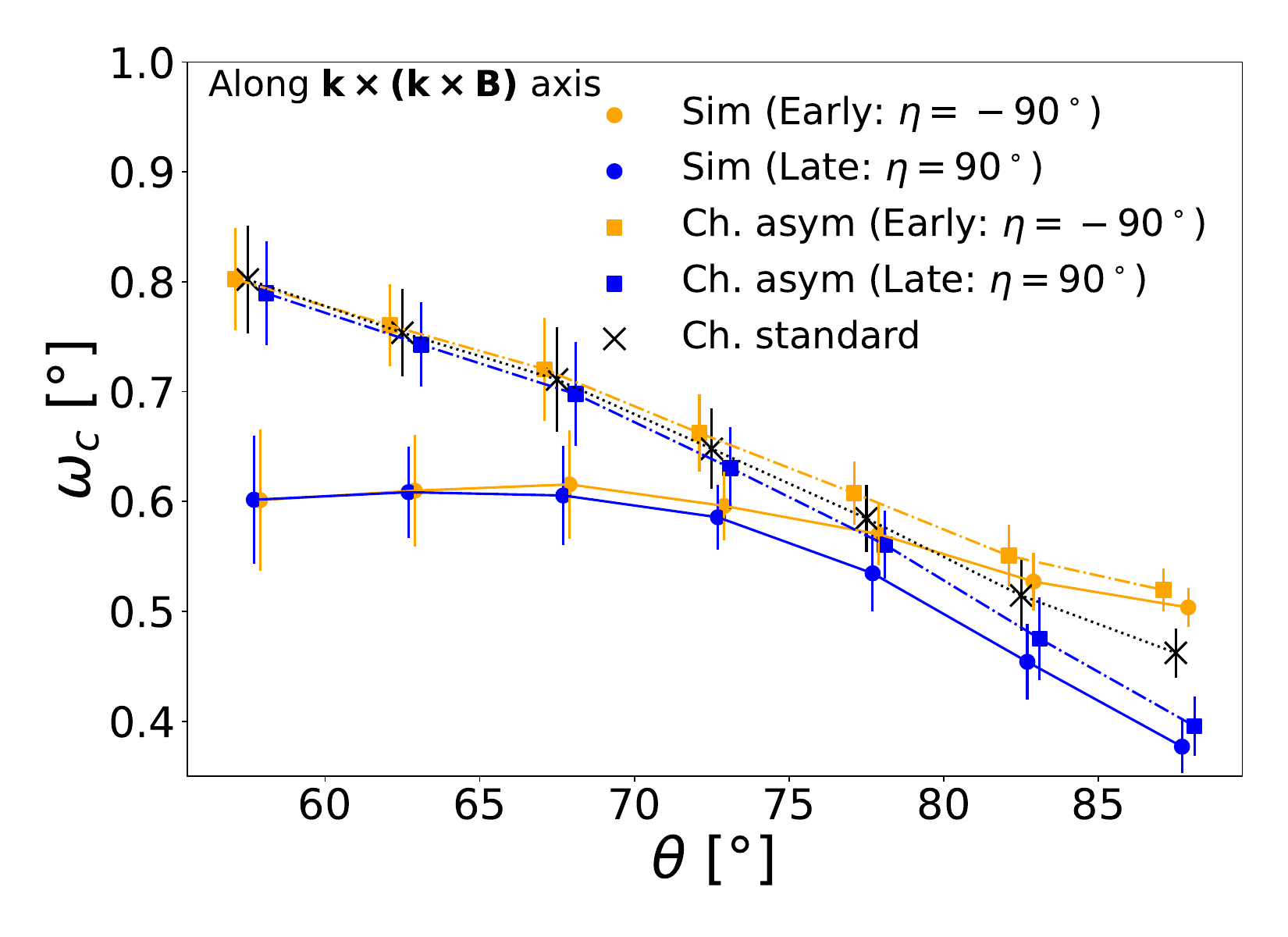}
         \includegraphics[width=0.49\columnwidth]{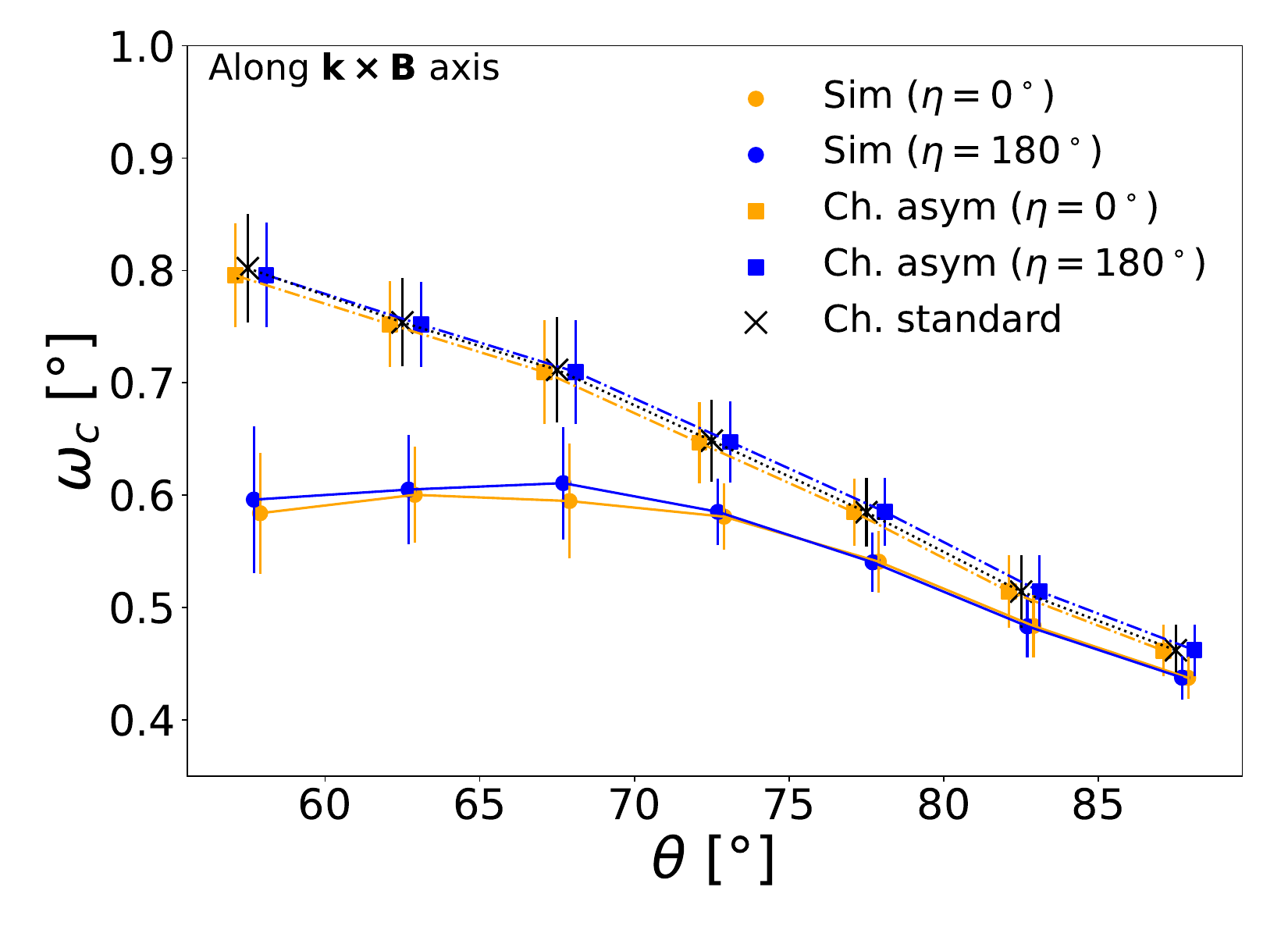}
     \caption{Evolution of the mean angular position of the amplitude maxima for simulated data with zenith angle $\theta$, along the $\mathbf{k} \times (\mathbf{k} \times \mathbf{B})$ ({\it left}) and $\mathbf{k} \times \mathbf{B}$ ({\it right}) axis (dots). Also shown are the Cherenkov angle values for the asymmetry toy model (squares) and for the standard Cherenkov computation (crosses). The reference point for angle computation is taken at the reconstructed $X_{e}$ position. In the plot along $\mathbf{k} \times (\mathbf{k} \times \mathbf{B})$ ({\it left}) early antennas ($\eta = -90\degree$) are shown in orange, late ($\eta = +90\degree$) in blue. 
     In both plots the error bars represent the standard deviation of the distributions. For each zenith angle, the points are artificially shifted for better visibility.
     }
      \label{fig:cherenkovangle}
\end{figure}

We perform a more systematic evaluation of the Cherenkov asymmetry model by comparing the angular position of the amplitude maximum, averaged over the full simulation set, to the angular Cherenkov angle computed with the standard and asymmetry models. The result is shown in Figure\,\ref{fig:cherenkovangle}. 

We will discuss first the results found for zenith angles larger than 70$\degree$\, along the $\mathbf{k} \times (\mathbf{k} \times \mathbf{B}$) axis. We note that the angular position of the amplitude maxima decreases with increasing zenith angle, an effect expected as inclined air showers develop higher in the atmosphere, i.e. at lower air density, where the refractive index is lower, resulting in a narrower Cherenkov angle. The standard Cherenkov model follows the same trend (with a small offset), and the Cherenkov asymmetry model is equivalent to it. 

Yet along the $\mathbf{k} \times (\mathbf{k} \times \mathbf{B}$) axis, the asymmetry between early and late antennas appears clearly for these inclined trajectories and is well reproduced by the Cherenkov asymmetry toy model. As mentioned already, the standard model falls at an intermediate position. 

Regardless, for both plots, an offset between the asymmetry models and the simulated data is observed. This offset is between 0.02$\degree$ and 0.05$\degree$, that we consider acceptable given the limitations of the model (see next section), and our target for angular resolution around 0.1$\degree$.

For zenith angles below 70$\degree$, the asymmetry along the $\mathbf{k} \times (\mathbf{k} \times \mathbf{B}$) axis between positive and negative $\eta$ values becomes negligible. This is supported by both the simulations and the Cherenkov asymmetry model, as the differences in optical paths for early and late antennas become negligible. Yet while the Cherenkov angle computed with both standard and asymmetry models decreases nearly linearly with increasing zenith angle, in simulations the position of the maximum amplitude remains constant around 0.6$\degree$, yielding a significant offset of 0.2$\degree$ for $\theta = 60\degree$. Possible causes for this discrepancy between simulations and models are discussed in the following section.

\subsection{Systematic effects on the amplitude profile maxima}
\label{sec:systematic_effects}
We have identified two systematic effects impacting significantly the determination of the angular position of the amplitude profile maxima.

\subsubsection{Radio signal frequency range}
\label{sec:frequency_effect}
\begin{figure}[ht!]
     \centering
     \includegraphics[width=0.48\columnwidth]{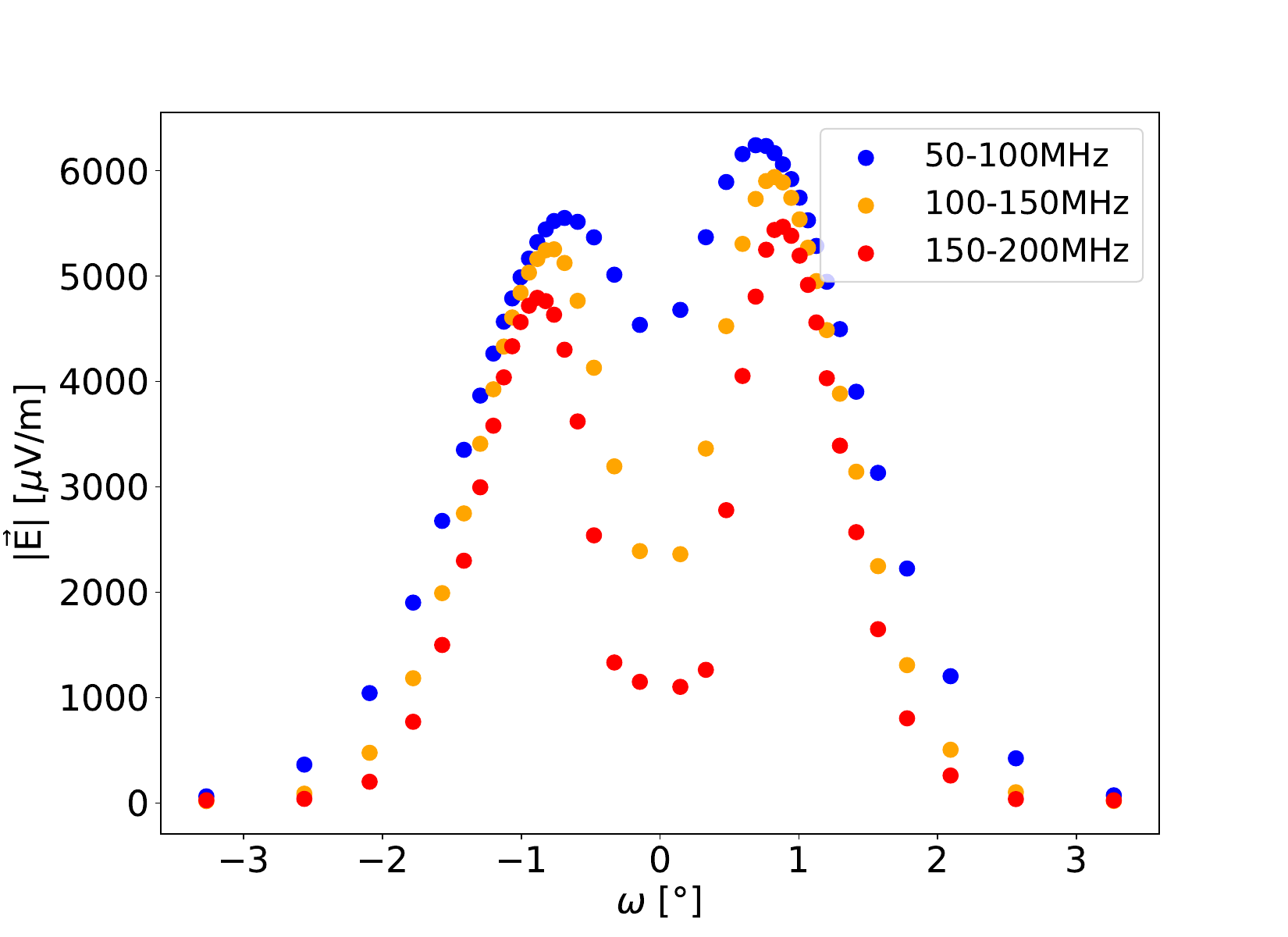}
     \includegraphics[width=0.48\columnwidth]{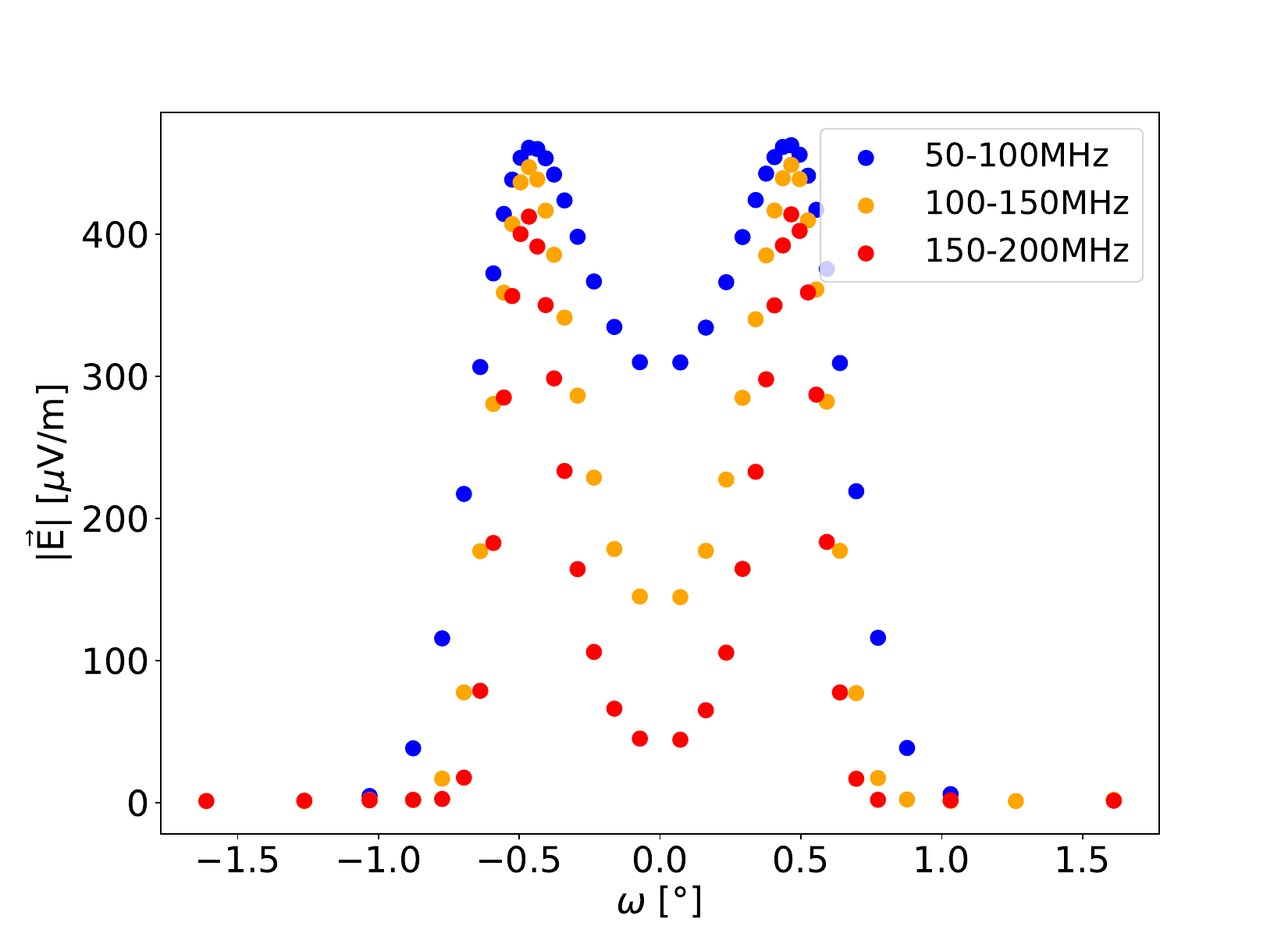}
      \caption{Illustration of the frequency effect on the angular position of the peak amplitude. The electric field is filtered in three distinct frequency ranges and the signal amplitude distribution is displayed along the $\mathbf{k} \times \mathbf{B}$ axis for (\textit{left}) $\theta = 57.0\degree$ and (\textit{right}) $\theta = 87.1\degree$. The $\omega$ angle is computed from the true $X_{\rm max}$ position as no reconstruction is performed here.}
    \label{fig:frequency}
\end{figure}

To evaluate how the position of the amplitude maximum depends on the radio signal frequency range, we filter the signals from the simulation sets in the frequency ranges 50-100\,MHz, 100-150\,MHz, 150-200\,MHz,
before determining their peak amplitude. Figure~\ref{fig:frequency} shows the resulting profiles for an inclined ($\theta$ = 87.1$\degree$) and less inclined ($\theta$ = 57.0$\degree$) shower.

While for inclined showers the position of the Cherenkov angle is very similar for the three frequency ranges considered, an inward shift is observed for decreasing frequencies in the case of less inclined showers. This effect makes the modelling of the Cherenkov angle more difficult for low inclinations. 

\subsubsection{Computation of the \texorpdfstring{$\omega$}{w} angle}
\label{sec:point_source_position_effect}
As mentioned in section \ref{section:coordinates}, the amplitude distribution is described with angular coordinates, with the emission position $X_{e}$ taken as the origin for the $\omega$ angle computation. 
Hence, for a given shower geometry, 
the further away $X_{e}$ is reconstructed, the smaller the $\omega$ angles will be for a same position at ground.

\begin{figure}[!ht]
         \includegraphics[width=0.5\columnwidth]{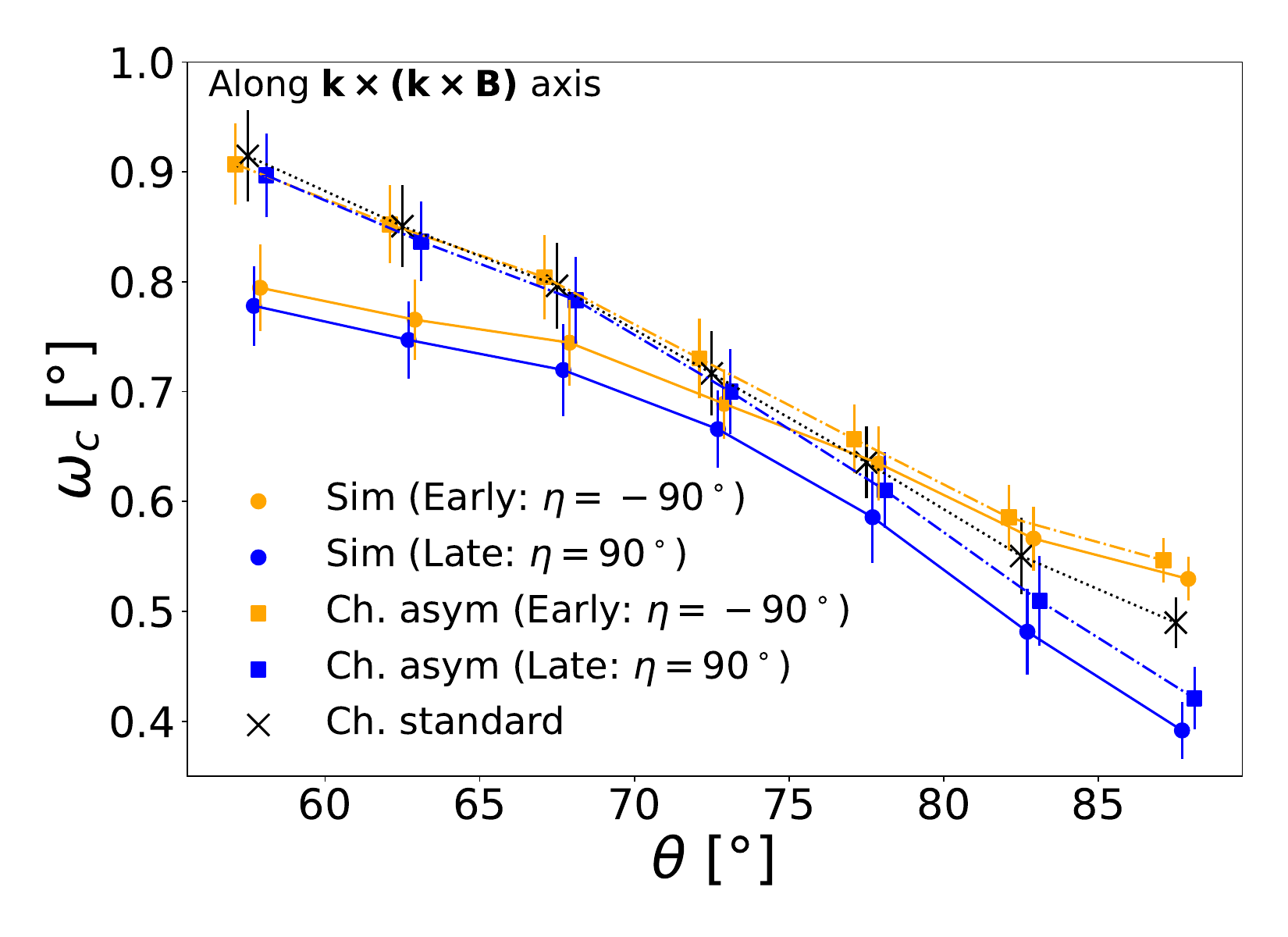}
         \includegraphics[width=0.5\columnwidth]{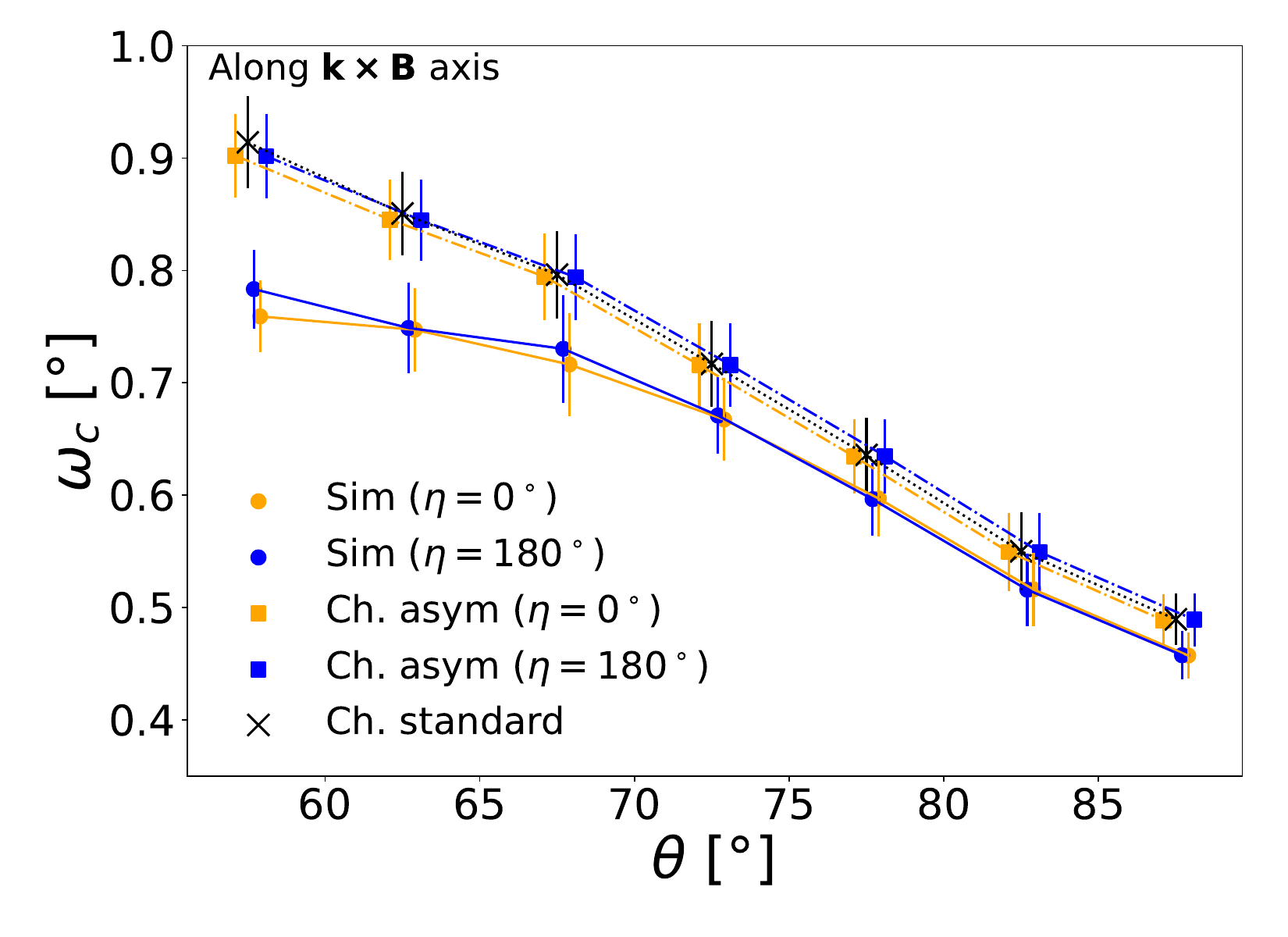}
     \caption{Evolution of the Cherenkov angle $\omega_c$, computed from the true $X_{\rm max}$ position, with zenith angle $\theta$, along the $\mathbf{k} \times (\mathbf{k} \times \mathbf{B})$ ({\it left}) and $\mathbf{k} \times \mathbf{B}$ ({\it right}) axis. The large dots represent the values obtained from the simulations, the squares are the values derived from our model, and the black crosses show the standard computation. In orange, ({\it right}) for $\eta = -90\degree$ (early antennas) and ({\it left}) for $\eta = 0\degree$. In blue, ({\it right}) for $\eta = 90\degree$ (late antennas) and ({\it left}) for $\eta = 180\degree$. The error bars represent the standard $\sigma$ value.}
      \label{fig:cherenkovangle_Xmax}
\end{figure}

To illustrate this effect, Fig.~\ref{fig:cherenkovangle_Xmax} represents the same data as in Fig. \ref{fig:cherenkovangle}, but with the Cherenkov angle $\omega_{c}$ computed from the true $X_{\rm max}$ position instead of $X_e$.
For zenith angle larger than $\theta = 70\degree$, the simulated data and the models give similar results to those observed on Figure~\ref{fig:cherenkovangle}, because $X_e$ is very close to $X_{\rm max}$. This result is consistent with the approximation of a point-like radio source close to shower maximum for inclined EAS. 

Yet for air showers with $\theta \leq 70\degree$, the peak amplitudes are shifted by $\sim0.1-0.2\degree$ compared to those observed in Figure~\ref{fig:cherenkovangle}. This offset is due to a source position reconstructed further away from $X_{\rm max}$ for $\theta \leq 70\degree$ as illustrated in Figure~\ref{fig:SWF_stats} from the appendix. This shows the limits of the point source hypothesis for vertical showers where a hyperbole describes more accurately the wavefront, as observed by LOPES~\cite{ Corstanje_2015} or LOFAR~\cite{Apel_2014}. 

\subsection{Refractive index computation}
We evaluate the impact of the true refractive index value on the Cherenkov asymmetry model by applying an additional $10\%$ offset to the final value of $n_{\rm eff}-1$.
This offset represents a safe upper limit to natural variations in  refractive index coming from changes in atmospheric temperature, humidity, and air pressure~\cite{CORSTANJE_2017}. This results in a relative error on the Cherenkov angle (from the previous one computed using the true $n_{\rm eff}$ from ZHAireS simulations) of less than $5\%$, allowing us to conclude that the refractive index has a negligible influence on the position of the $\omega_c$ angle compared to the effects mentioned in the two previous subsections.

\vspace{1cm}
We have seen in the previous sub-section that the description of the EAS amplitude profiles by an Amplitude Distribution Function highly depends on the radio frequency range considered and on a good spherical reconstruction of the radio wavefront. This evidences that the model proposed in this article and centered on the ADF provides only a limited, model-dependent, and empirical description of the EAS radio profile. Yet it appears to be remarkably robust for showers with zenith angles larger than 70$\degree$, producing, in particular, a precise estimate of the positions of the amplitude maxima and their asymmetry. 
For zenith angles below 70$\degree$, the asymmetry effect becomes negligible and the angle of maximum amplitude remains constant around $0.6\degree$, in part as a compensation for the non-sphericity of the shower front. Hence, in this configuration, the parameter $\omega_c$ can be chosen as the minimum between $0.6\degree$ and the value predicted by the asymmetry model, as shown in Figure~\ref{fig:cherenkov_assym_model} (\textit{Right}). 

Fixing the value of $\omega_c$ provides a powerful lever arm to reconstruct precisely the direction of origin of the EAS, as will be shown in the next section.

\section{Direction reconstruction using the ADF}
\label{section:reconstruction}
In the previous sections, we have seen that the radio amplitude pattern of air showers with zenith angle larger than 60$\degree$ can be successfully described by an analytical function --the Angular Distribution Function-- depending solely on the shower geometry (defined by the direction of propagation $\mathbf{k}$ and one emission point $X_{e}$, describing together the shower axis) and two additional parameters (a scaling factor and the width of a Lorentzian). 

In this section, we will outline the general reconstruction procedure and its respective performance obtained on the direction reconstruction using a set of realistic simulations.

\subsection{Method}
\label{sec:method}
The complete reconstruction pipeline follows three steps. Each of these steps is processed distinctly and the results of one step are used as input parameters to the next.

First, the arrival times of the signals at the antenna position are fitted with a plane wave. This is done in our treatment using an analytical method presented in~\cite{ferrière2024} and outputs a reference direction ($\theta_0$, $\phi_0$) for the shower direction of origin. This step is not mandatory in principle, but it reduces the parameter space considered for the spherical and ADF fits in the next steps and thus accelerates their convergence.

Then, the arrival times are adjusted with a spherical curvature model to determine the position of the point-like source $X_{e}$, which is searched in a 3D volume around the direction of $\qty(\theta_0, \phi_0)$ (see \ref{annex:mini} for parameter details). For each tested location, the spherical wave is computed taking into account the effective refractive index along the optical path of each antenna. Together, $X_e$ and $\qty(\theta_0, \phi_0)$ define an initial shower axis, whose direction is refined in the next step while $X_e$ position remains fixed to the result of the spherical fit.

Finally, the amplitude profile of the radiation footprint is fitted in the shower angular plane with the ADF by minimizing the residual function 
\begin{eqnarray}
\label{eq:minimisation}
R_{\theta, \phi} = \sum_{i=1}^{N_{\rm antennas}} \left[ A_{i} - f_{i}^{\rm ADF}(A, \theta, \phi,\delta \omega;  x_{e}, y_{e}, z_{e}) \right]^2
\end{eqnarray}
where $f_{i}^{\rm ADF}(\theta, \phi, \delta \omega, A; x_{e}, y_{e}, z_{e})$ is given by Eq.~\ref{eq:adf} with all the angles computed with respect to the emission point position ($x_{e}, y_{e}, z_{e}$) and $A_{i}$ the peak amplitude at each antenna computed as described in Section~\ref{section:simus}.

Four free parameters are adjusted through the minimization of $R_{\theta, \phi}$: the angles $\theta$ and $\phi$ defining the shower direction,  the scaling factor $A$ and the Lorentzian width $\delta \omega$. The scaling factor $A$  and the Lorentzian width $\delta \omega$ are the only free parameters appearing explicitly in the ADF formula (see Eqs. \ref{eq:adf} and \ref{eq:cher}), yet the antenna coordinate $\omega$ relate to the shower direction of propagation ($\theta$, $\phi$), as well as the emission point position ($x_{e}, y_{e}, z_{e}$), through Eq.\,\ref{eq:pos}, hence $R$ depends on $\theta$ and $\phi$.

The antenna coordinates in the ground frame, their associated trigger times (for the reconstruction of the initial direction and the point source position) and the peak amplitudes are the only input to the reconstruction procedure. 

Finally, we will point that the uncertainty on the value obtained for the adjusted parameters is presently not computed in the process. This will done in a further stage of our work.

\subsection{Simulation set}
\label{section:simus_realistic}
\begin{figure}[!ht]
     \centering
    \includegraphics[width=0.51\columnwidth]{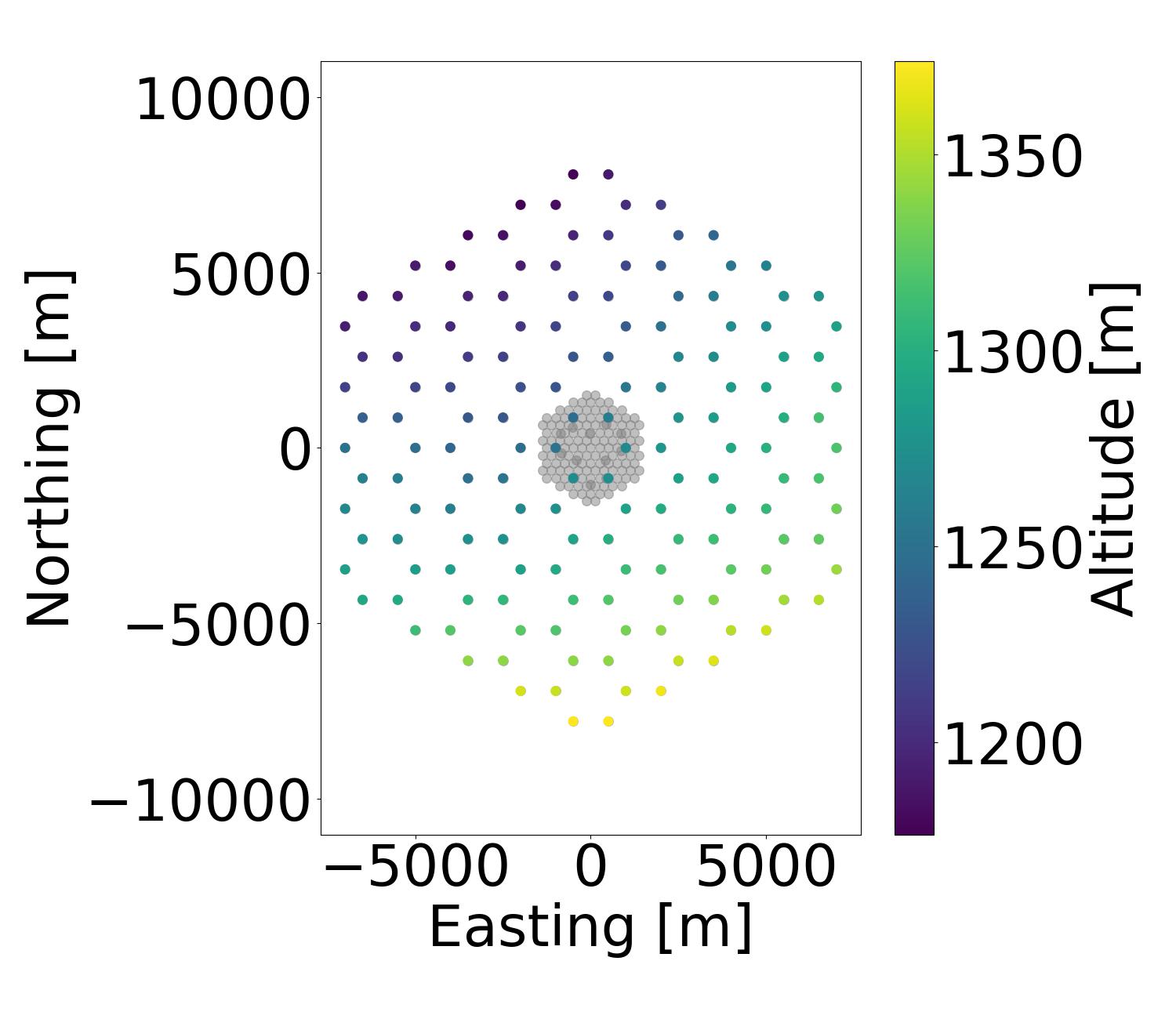}
    \includegraphics[width=0.45\columnwidth,trim=55 170 60 0, clip]{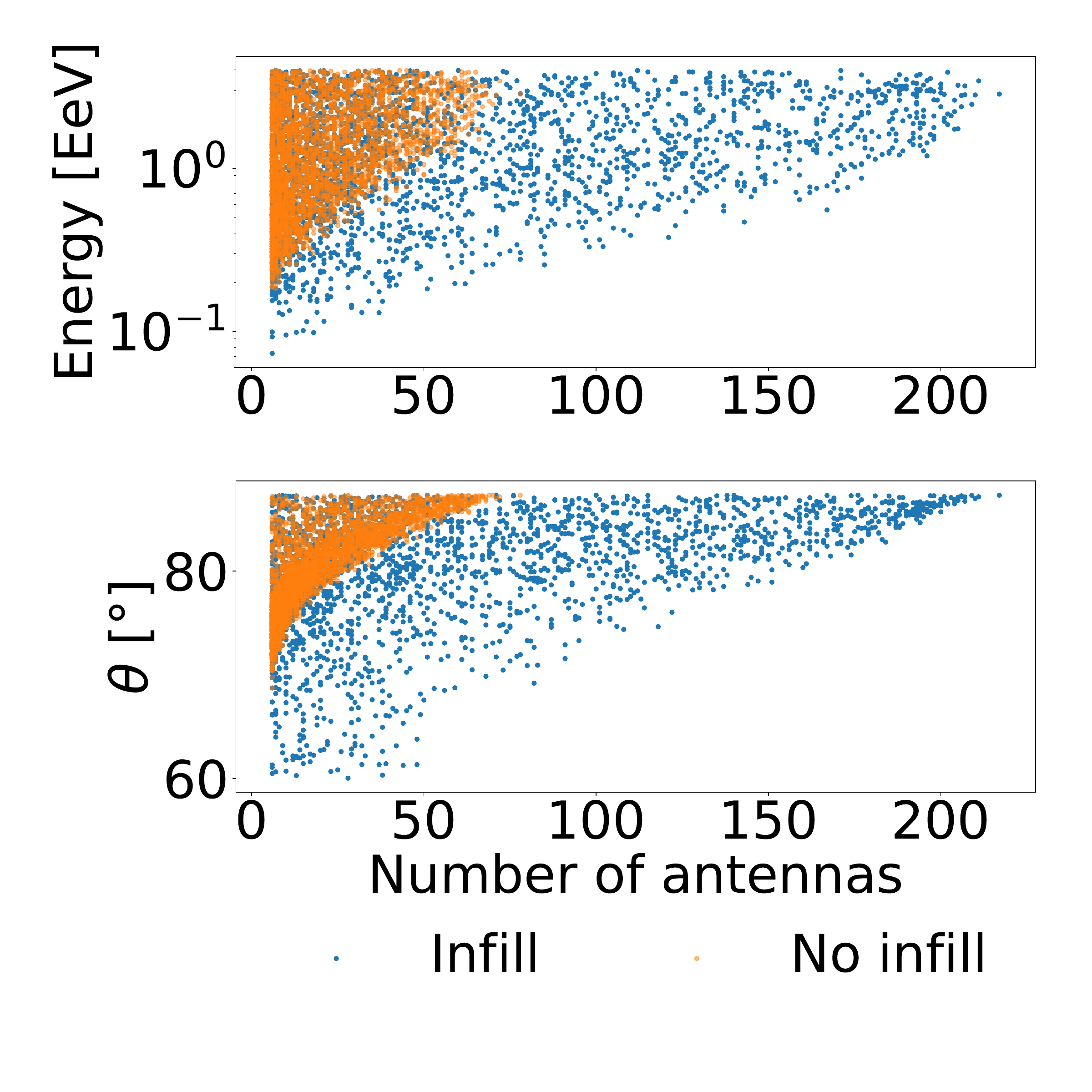}
      \caption{(\textit{Left}) Hexagonal antenna layout taking into account the topography at the GP300 site \cite{Martineau-Huynh_hdr2021}. The infill is represented in grey. (\textit{Right}) Number of triggered antennas per event as a function of energy (\textit{top}) and zenith angle (\textit{bottom}), including only events with $N_{\rm ants} \geq 6$. Results are shown for the full array with infill (blue) and without infill (orange).}
    \label{fig:layout}
\end{figure}

The performance for direction reconstruction of the method described in the previous section is evaluated using a realistic layout consisting of a hexagonal pattern with 150 antennas spaced 1 km apart. Additionally, we consider an infill with 134 extra antennas spaced 250\,m apart (see Figure~\ref{fig:layout} left). The altitude of the antennas are determined by the topography of the GP300 site~\cite{chiche_2024_gp300}. In this section, we analyze both layouts: one with the infill added to the hexagonal pattern previously mentioned, and one without. The layout with infill extends the detectable zenith angle range down to approximately 60$\degree$ (see Figure~\ref{fig:layout}, right) and slightly broadens the energy range. 

The simulation set comprises $\sim 10^4$ EAS induced by protons and iron nuclei. They are simulated with ZHAireS with energies ranging between $10^{16.5}\,\rm eV$ and $10^{18}\,\rm eV$ with logarithmic bins, the azimuth angle distribution is uniformly distributed between $0\degree$ and $360\degree$ and the zenith angle distribution varies between $60\degree$ and $87.3\degree$ over logarithmic bins of $1/\rm cos(\theta)$. The shower core positions are randomly drawn inside the antenna array. Therefore only core-contained events are simulated here.

In realistic conditions, Galactic emissions and other background sources introduce random noise into the radio signal, affecting the accuracy of the peak time and amplitude. To account for this, a random Gaussian noise signal with a standard deviation of $(\sigma_{\rm Gal}/\sqrt{3} = 13\; \mu \rm{V/m})$ is added to the signal for each polarization. This value represents the average level of electromagnetic radiation induced by the Galaxy within the 50--200 MHz frequency range along one polarization, based on the calculations presented in~\cite{decoene_2021}.
The simulated electric field traces are then processed as described in Section~\ref{section:simus}: the norm of the electric field is computed as the quadratic sum of the 3 components, and the maximum of its Hilbert envelop is taken as the electric field amplitude.
A shower is eventually considered as detected if at least 5 antennas exhibit an amplitude larger than $5\sigma_{\rm Gal} = 110\,\rm \mu V/m$.  About 4000 events from the simulation set pass this cut. 

Calibration uncertainties are also taken into account by smearing the peak amplitudes with a random Gaussian error with standard deviation $\sigma_{\rm A} = 7.5\%$\,\cite{PierreAuger:2017xgp}. GPS jitter time is accounted for by randomizing the trigger times with a Gaussian distribution of $\sigma_{\rm t} = 5\,\rm ns$.

\subsection{Performance}
\label{perfs_ang}
The accuracy of the ADF reconstruction of the arrival direction is estimated by computing the angular distance $\psi$ between the true shower direction and the reconstructed one with:
\begin{eqnarray}
\label{eq:angular_distance}
cos(\psi) = \rm cos(\theta_{\rm rec}) cos(\theta_{\rm true}) + cos(\phi_{\rm rec} - \phi_{\rm true}) sin(\theta_{\rm true}) sin(\theta_{\rm rec}) 
\end{eqnarray}
where $\theta_{\rm rec}$ and $\phi_{\rm rec}$ are respectively the reconstructed zenith and azimuth --corresponding to the minimal $\Delta$ value, see section \ref{sec:method}--, $\theta_{\rm true}$ and $\phi_{\rm true}$ the true zenith and azimuth. 

\begin{figure}[!ht]
     \centering
    \includegraphics[width=0.49\columnwidth,trim=10 25 20 0, clip]{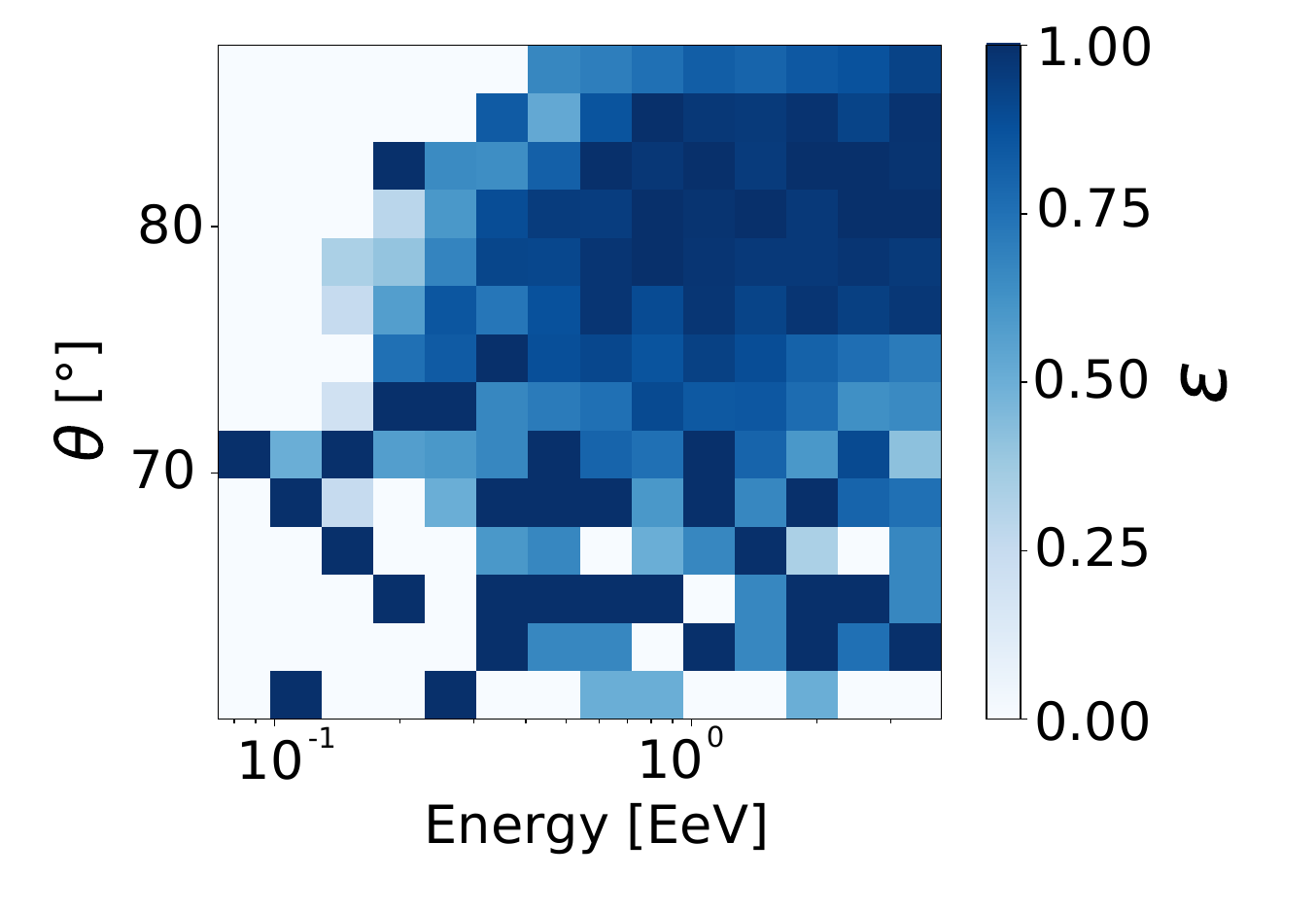}
    \includegraphics[width=0.5\columnwidth,trim=10 80 140 5, clip]{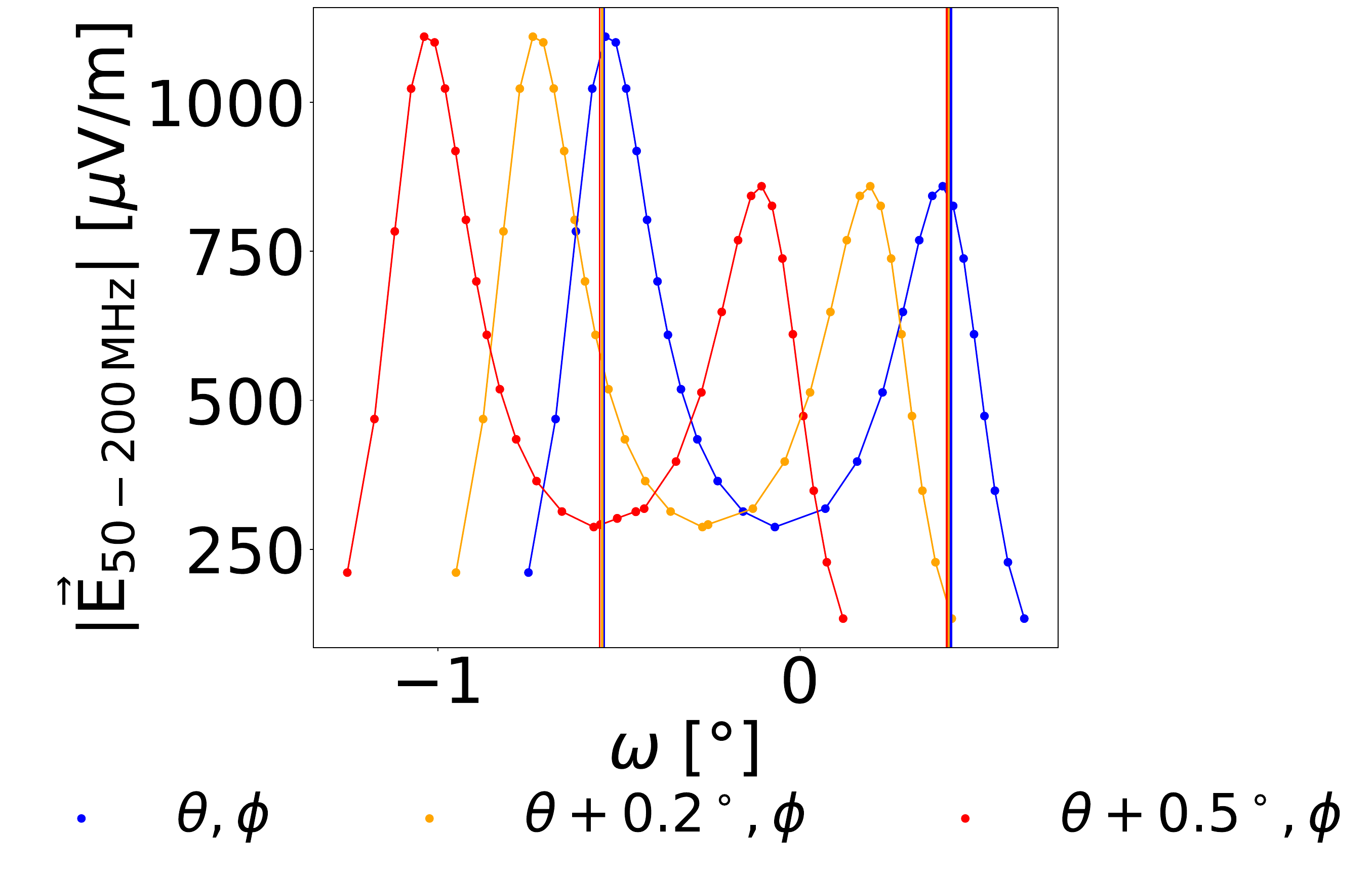}
      \caption{(\textit{Left}) Efficiency $\epsilon$ as a function of primary energy and zenith angle for the set of simulations with infill. The efficiency is defined as the ratio between the number of reconstructed events and the total number of triggered events. (\textit{Right}) Illustration of the ADF fit mechanism: signal amplitude distribution along the $\bf{k} \times (\bf{k} \times \bf{B})$ axis for one event induced by a proton with primary energy $E_{\rm primary} = 3.98\,\rm  EeV$ with $\theta = 86.5\degree$ and $\phi = 0\degree$. 
      Three different configurations are considered: $\omega$ and $\eta$ computed from the true direction $(\theta, \phi)$ (blue), $(\theta+0.2\degree, \phi)$ (orange) and $(\theta+0.5\degree, \phi)$ (red). The Cherenkov angle positions computed from the Cherenkov asymmetry model for these 3 directions are also indicated by the vertical lines. Note that in the 3 cases, $X_e=X_{max}$ is considered. As anticipated, the positions are nearly identical across the three configurations.}
    \label{fig:adf_badprofile}
\end{figure}

Only events with at least 6 triggered antennas ($N_{\rm ants} \geq 6$) are finally selected. This represents $99 \%$ of triggered showers in our simulation set with and without infill. The ADF fit converges for $88\%$ ($90\%$) of these selected events with (without) infill.  Figure~\ref{fig:adf_badprofile} (left) displays the reconstruction efficiency $\epsilon$ as a function of zenith angle and energy, using the set of simulations with infill. The efficiency is defined as the ratio between the number of events successfully reconstructed —i.e. events with $N_{\rm ants} \geq 6$ and a converging ADF fit —and the total number of triggered events. As expected, the reconstruction efficiency increases with both energy and inclination.

\begin{figure}[!ht]
     \centering
    \includegraphics[width=0.480\columnwidth,trim=10 0 20 0, clip]{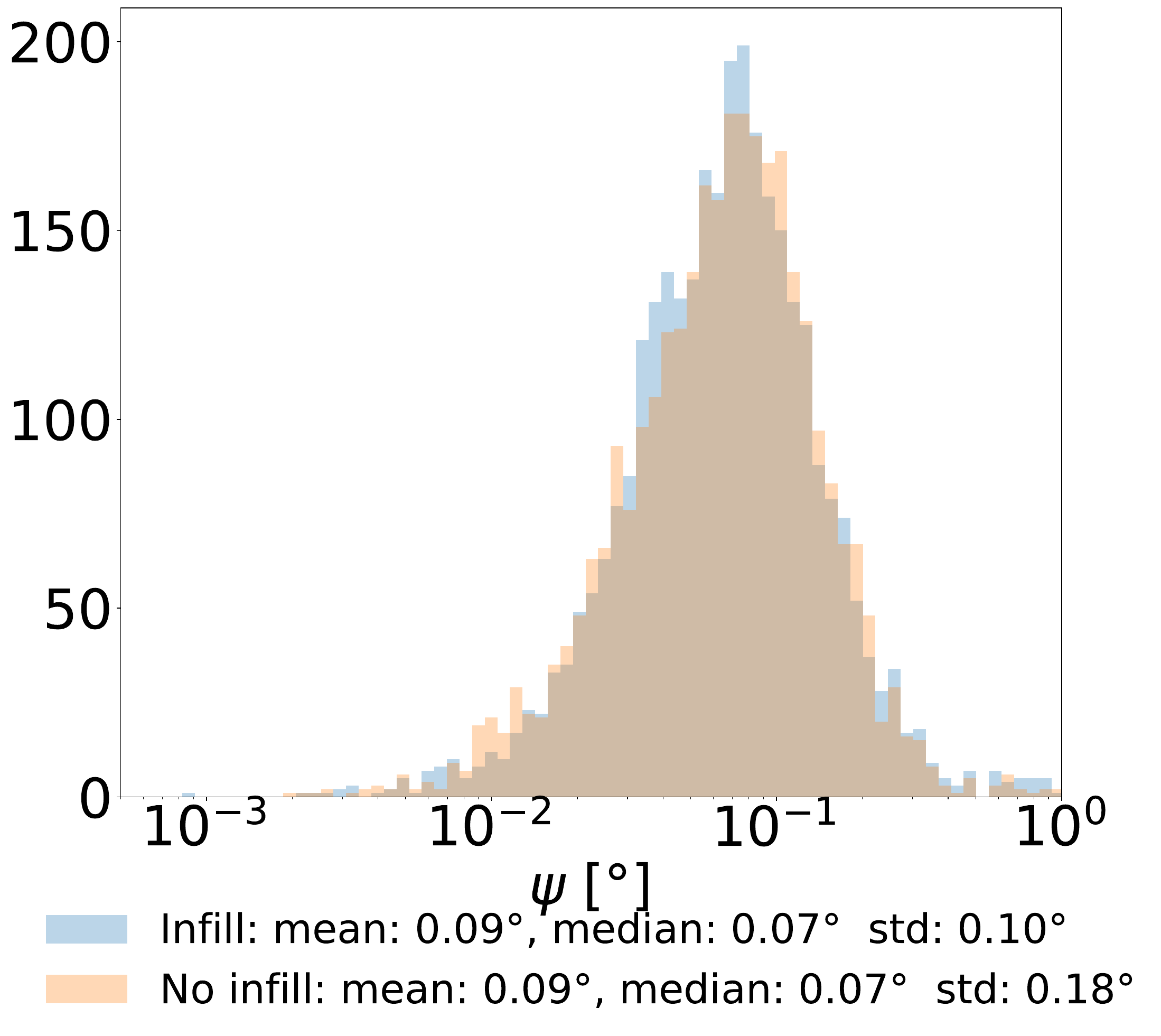}
\includegraphics[width=0.49\columnwidth]{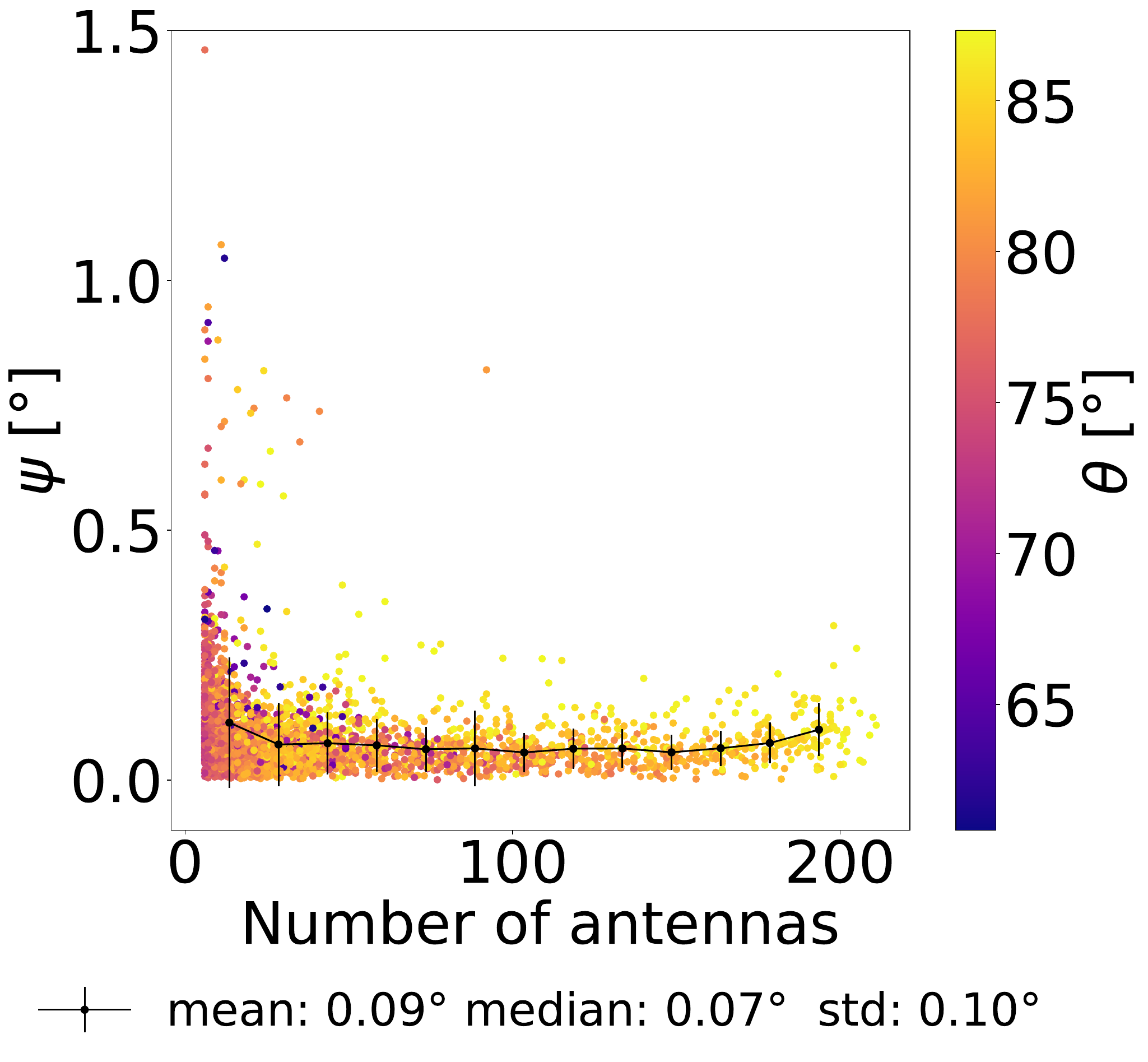}
    
      \caption{(\textit{Left}) Distribution of the angular distance $\psi$ between the true shower direction and the reconstructed for simulations on the hexagonal layout with (without) infill in blue (orange). (\textit{Right}) Angular distance $\psi$ as a function of number of antennas and zenith $\theta$ for the hexagonal layout with infill.}
    \label{fig:adf_histo}
\end{figure}

The angular resolution $\psi$ for these events is displayed in the left panel of Figure~\ref{fig:adf_histo}. Its median value is 3.6 arc-minutes ($0.07\degree$), with 80\% of events below $0.1\degree$.

\begin{figure}[!ht]
     \centering
     \includegraphics[width=0.49\columnwidth,trim=10 0 20 40, clip]{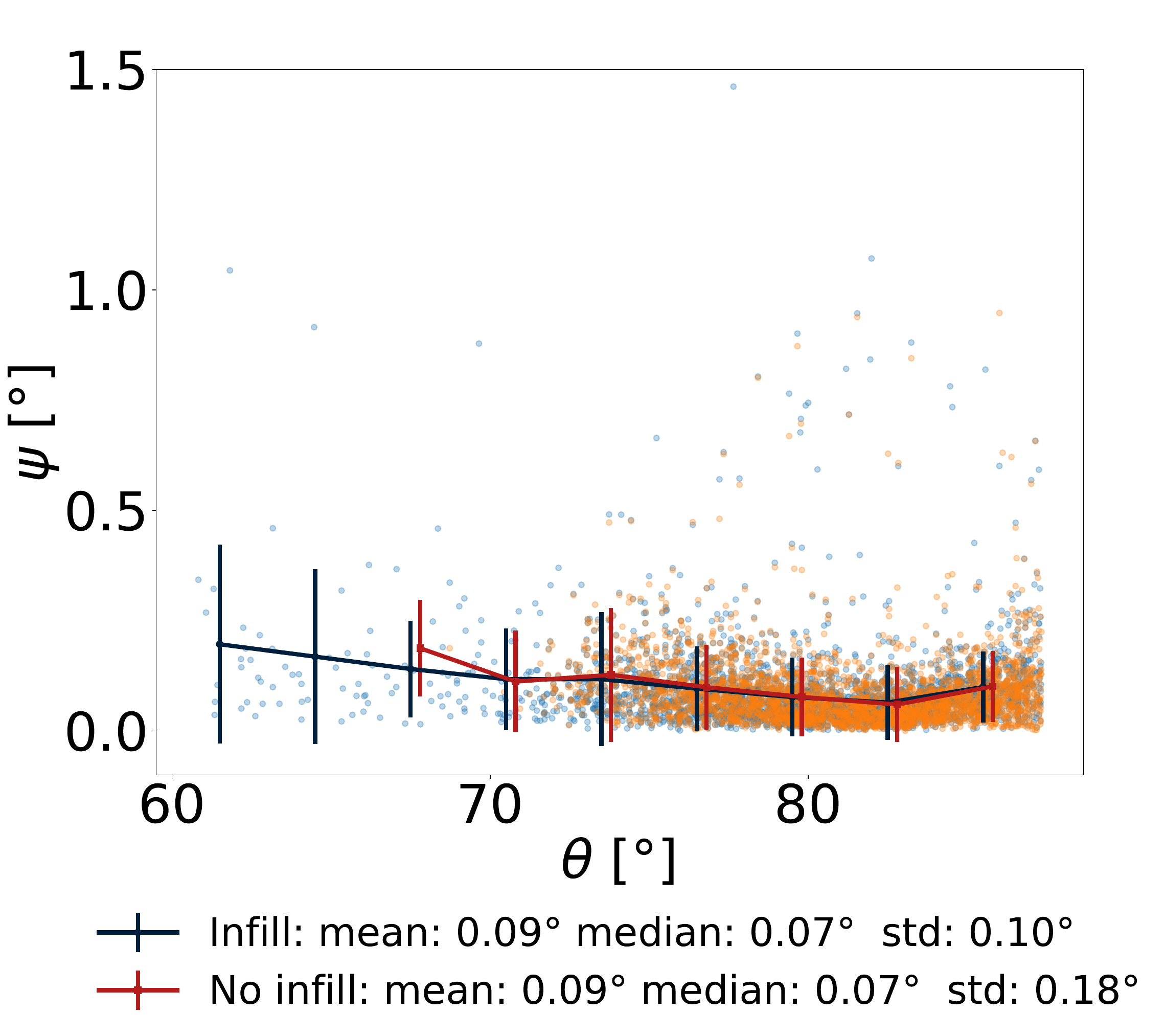}
     \includegraphics[width=0.49\columnwidth]{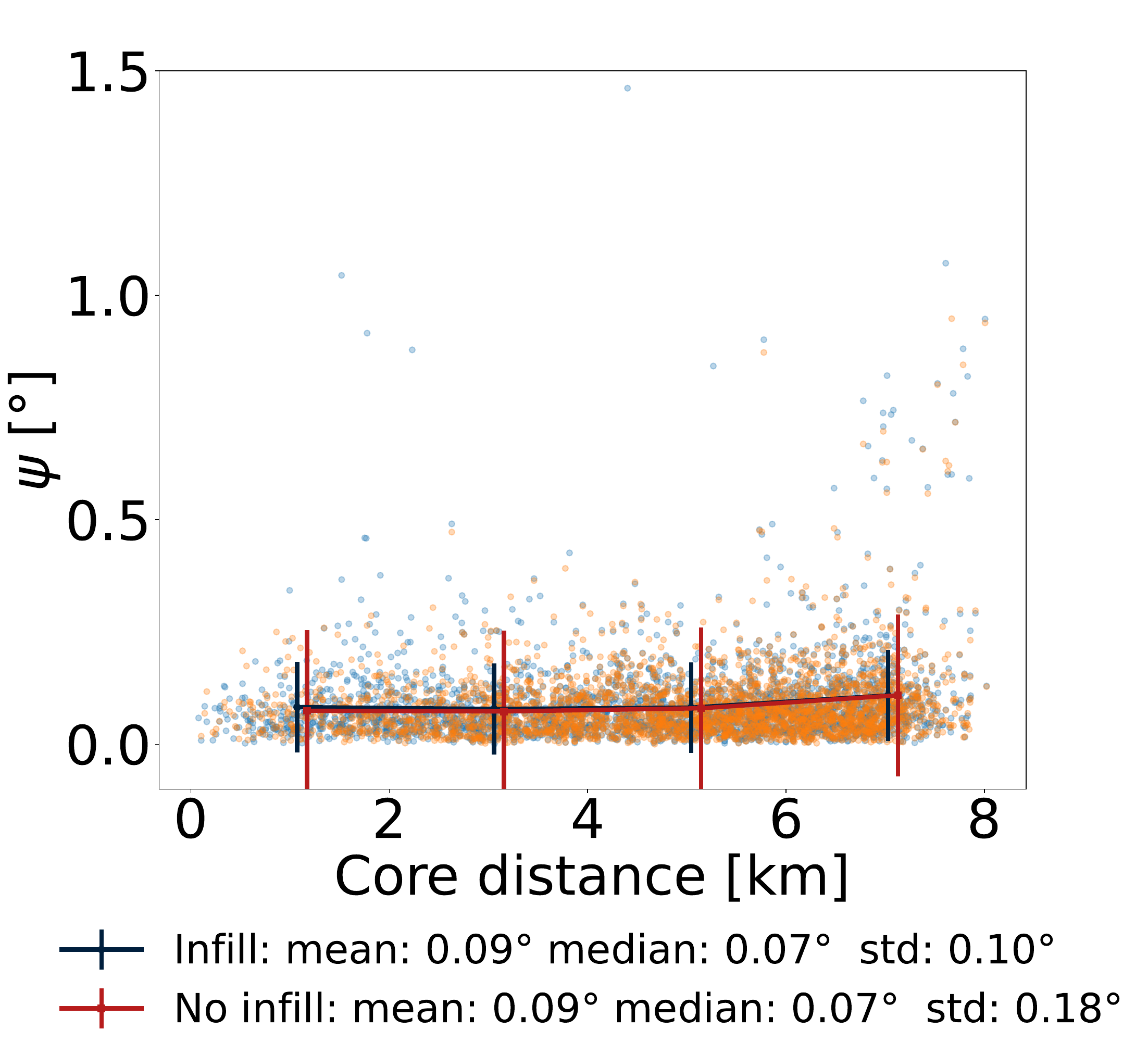}
      \caption{ Angular distance $\psi$ as a function of zenith angle $\theta$ (\textit{Left}) and core position (\textit{Right}) with respect to the center of the array. For both 
    panels, the reconstruction is performed on the complete hexagonal layout including the infill (blue) or without the infill (orange). The main array spans a hexagon with side length 8.1 km, while the infill array covers a smaller hexagon with side length 1.625 km.}
    \label{fig:adf}
\end{figure}

The angular error does not show significant variation with antenna number down to 15 units (see Fig. \ref{fig:adf_histo} right), shower core position, nor zenith angle down to $\sim60\degree$ (see Figure~\ref{fig:adf}). The latter can be explained by the fact that the position of the reconstructed source $X_e$ is offset from the true shower axis by a few tens of meters only~\cite{decoene_2023} (see~\ref{appendix:direction}) in the zenith angle range $60\degree-70\degree$, even if the point-source-like description is then not valid anymore. The good performances up to large core positions shows that the reconstruction remains reliable at the edge of the layout, where the radio footprint is only partially detected. 
 
A similar resolution is achieved on the direction of origin for the sparse array (see Figure \ref{fig:adf_histo} left), but the zenith range is then limited to $\theta \geq 68\degree$ in our simulation dataset because of trigger efficiency (see Figure\,\ref{fig:layout} right). 

Although the plane wave reconstruction already achieves a sub-degree angular resolution (see~\ref{appendix:direction}), the ADF fit improves these results by a factor $\sim$2, in particular for more vertical, small number of triggered antennas or for core positions on the edge of the layout. The latter aspect will be studied in a latter study with non-contained cores, where plane wave reconstruction is expected to be less efficient. 

We will conclude this section by pointing that the excellent reconstruction performances of the ADF are, in our understanding, mostly due to the fact that it relies on four adjustment parameters only, with the position of the amplitude maxima fixed. This induces a very high sensitivity on the direction, as an offset to true direction quickly induces a bad amplitude profile which cannot be adjusted by ADF, as illustrated in Figure \ref{fig:adf_badprofile} (right). This plot illustrates the leverage provided by the toy-model computation of the Cherenkov angle: even when derived from an incorrect direction ($\theta$, $\phi$), its value remains close to the true one --as long as the source position $X_e$ is correct--, while the signal pattern in the angular plane is significantly shifted. Hence fixing the parameter $\omega_c$ in Eq. \ref{eq:cher} to this computed value strongly constrains the ADF fit.

\section{Energy reconstruction using ADF}
\label{section:energy}
The ADF model enables the reconstruction of the electromagnetic energy $( E_{\text{em}})$ of the EAS using a fitted scaling factor $(A)$. This scaling factor is directly related to the amplitude of the total electric field $( \mathbf{E})$. In Sec. \ref{section:energy_method}, we present the method we developed for energy reconstruction using the ADF model. We illustrate this method using the star-shaped simulations described in Section~\ref{section:simus}. As for the study carried out in section \ref{perfs_ang}, our analysis is limited to EAS with 6 or more triggered antennas.
In Section \ref{section:energy_perfs}, we present the results of the energy reconstruction based on the simulation set presented in Section \ref{section:simus_realistic}.

\subsection{Method}
\label{section:energy_method}
\begin{figure}[!ht]
     \centering
    \includegraphics[width=0.49\columnwidth]{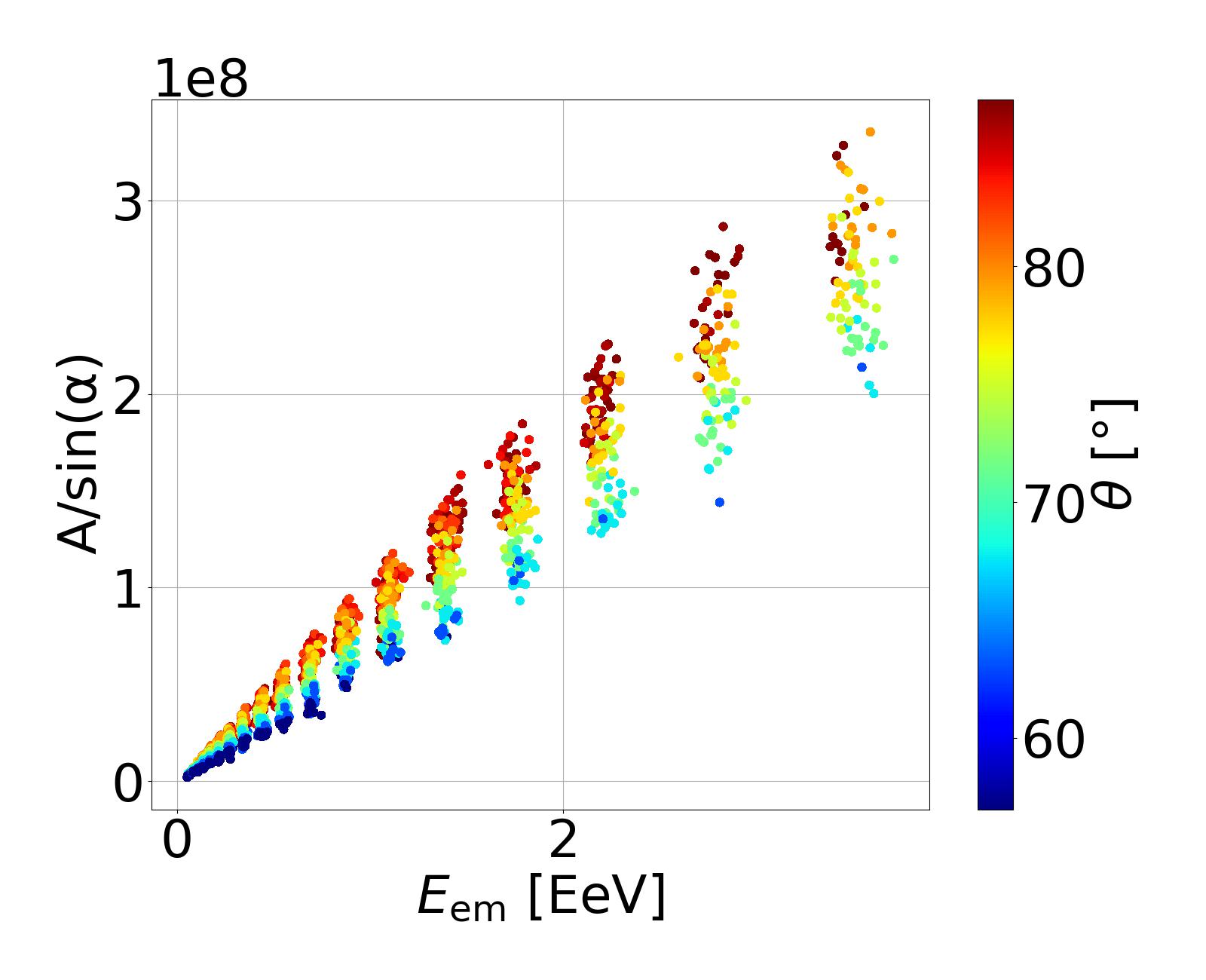}
    \includegraphics[width=0.49\columnwidth]{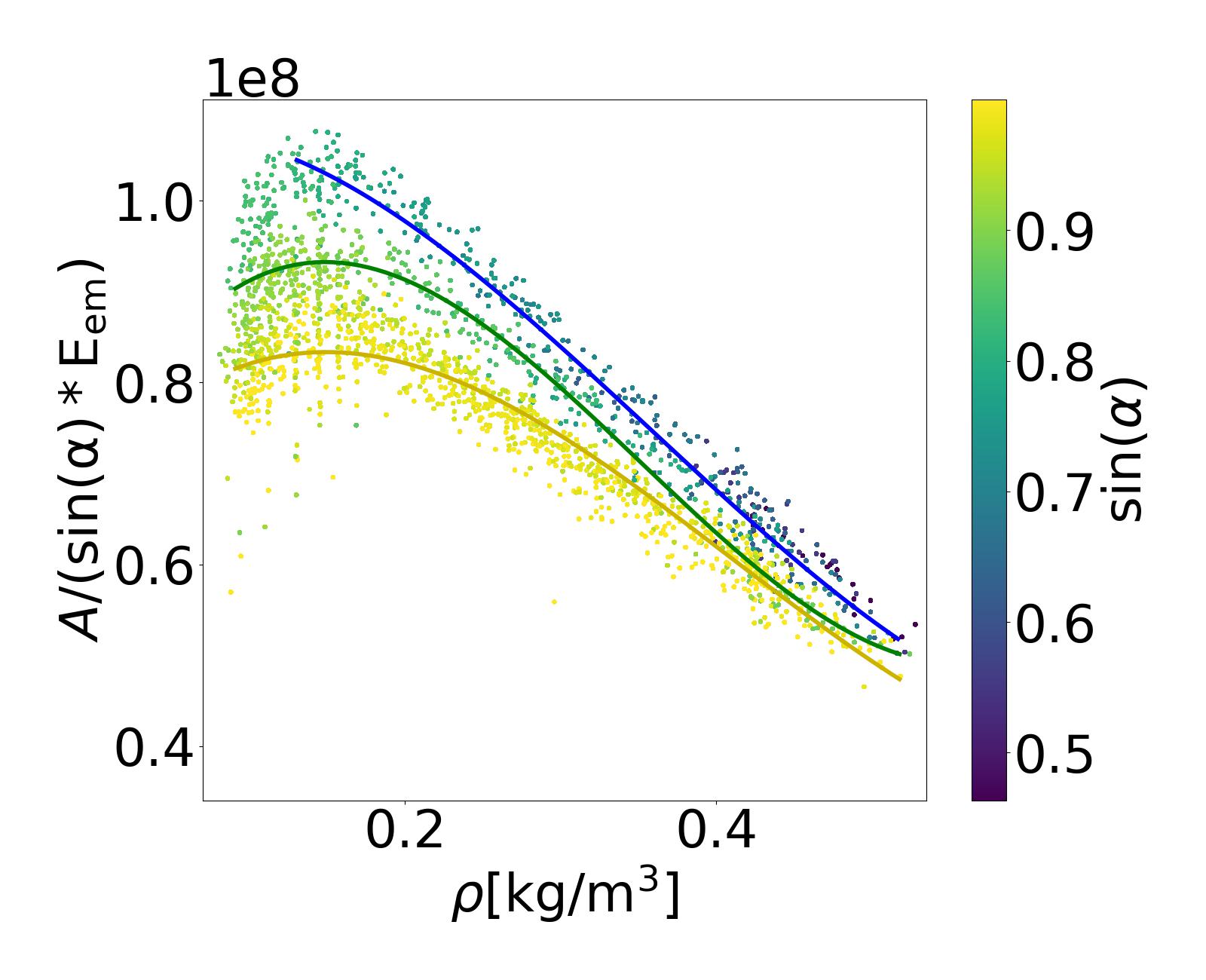}
      \caption{(\textit{Left}) Electromagnetic energy dependency on the corrected scaling factor $A/\sin{(\alpha)}$  for star-shape simulations. The color codes for the zenith angle of the EAS. A strong variation of the radio efficiency factor with the zenith angle is observed. (\textit{Right}) Air density dependency on the radio efficiency factor $A/(\sin{(\alpha)}E_{\rm em}$). A residual $\sin{(\alpha)}$ dependency is observed. The three curves represent the density correction factor $f$ applied on each range of $\rm sin(\alpha)$ (see text for details).}
    \label{fig:scalingfactor_energy}
\end{figure}

The geomagnetic effect---the dominant process for EAS radio emission---is induced by the deflection of electrons and positrons in opposite directions under the effect of the Lorentz force $\mathbf{F} = \pm e \mathbf{v} \times \mathbf{B}$, where v is the particle velocity. The amplitude of the radio signal therefore scales at first order as $e \norm{\mathbf{v}} B_{\rm{eff}}$, with $B_{\rm{eff}} = \sin{(\alpha)} \norm{\mathbf{B}}$ an effective magnetic field strength depending on $\alpha$, the geomagnetic angle between the Earth magnetic field and the shower axis. To account for this effect, the scaling factor $A$ is therefore first corrected by a factor $1/\sin{(\alpha)}$. The left panel of Figure~\ref{fig:scalingfactor_energy} shows the correlation between this corrected scaling factor and the electromagnetic energy $E_{\rm em}$ for our simulation set. Computation of $E_{\rm em}$ from the ZHAireS outputs is detailed in~\ref{sec:em_energy}. 

The distribution of $A/\sin{(\alpha)}$ depends linearly on $E_{\rm em}$ at first order, but a dispersion correlated with inclination is also visible. This is due to the fact that inclined showers develop higher in atmosphere, where the lower air density allows for a stronger geomagnetic effect, as already discussed in~\cite{Glaser_2016} and \cite{Chiche_2022}. 
To illustrate this, we display (right panel of Figure~\ref{fig:scalingfactor_energy}) the ratio $A/(\sin{(\alpha)}E_{\rm em})$---which we call ``radio efficiency''---as a function of the atmospheric density $\rho$, measured at the reconstructed $X_{e}$ position. In addition to the expected decrease of the radio efficiency with increasing air density, a dispersion related to $\sin(\alpha)$ is observed.
This was discussed in~\cite{chiche_2024} and \cite{chiche_2023} and we simply summarize the main argument here: for larger $\sin{(\alpha)}$, the stronger effective magnetic field $B_{\rm eff}$ experienced by particles in the air shower induces a larger drift for electrons and positrons in the air shower, resulting in a larger lateral extent of the shower~\cite{guelfand_2024}. This implies that coherence of the electromagnetic emission is lost for high frequencies, thus reducing the total strength of the EAS electromagnetic emission over the full 50-200\,MHz range. 
Finally, the same figure shows a decrease in radio efficiency at low atmospheric density, corresponding to high inclination angles. A loss of coherence may explain this reduction in efficiency; in very inclined EAS, particles experience more significant deflection due to reduced Coulomb scattering with air molecules. 

To summarize, an energy estimator can be built by applying to the scaling factor $A$ introduced in Eq. \ref{eq:adf} i) a coefficient $\sin(\alpha)$ accounting for the effect of shower geometry on the efficiency of the geomagnetic effect, ii) a coefficient $f(\rho, \sin(\alpha))$ depending primarily on the air density $\rho$ at the emission point, iii) but also on $\sin(\alpha)$ to account for secondary coherence effects. 

Our energy estimator can thus be written as\,:
\begin{eqnarray}
\label{eq:energy_estimator}    
E^{*}_{\rm em} = \frac{A}{{\sin{(\alpha)}} f(\rho, \sin{(\alpha)})}
\end{eqnarray}

In the following, the star-shape simulation set is randomly split into two subsets of equal size. The first one is used to determine the coefficient $f(\rho, \sin{(\alpha)})$ (in $\rm \mu V m^{-1} EeV^{-1}$). This is done by describing the distribution displayed in the right panel of Figure\,\ref{fig:scalingfactor_energy} with a polynomial regression of degree 3 between $\rm sin(\alpha)$, $\rho$, and the radio efficiency factor. It is worth noting here that this parametrization depends on the geomagnetic field value and the filtered frequency range.

\begin{figure}[!tb]
     \centering
     \includegraphics[height=0.5\columnwidth,trim=15 20 15 0, clip]{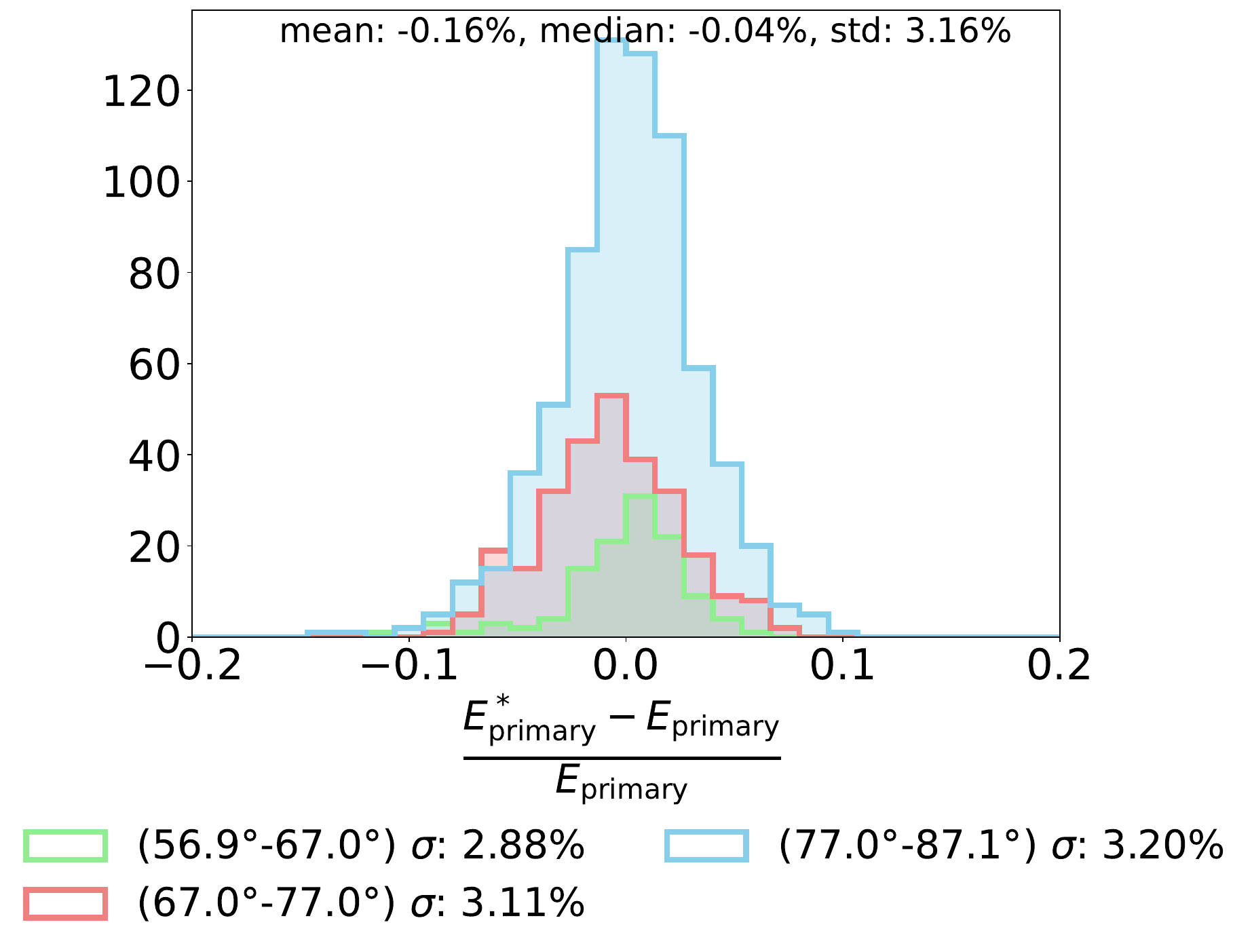}
      \caption{Electromagnetic energy resolution for different zenith ranges on starshape simulations. This corresponds to the intrinsic resolution achievable with the method.}
    \label{fig:resolution}
\end{figure}

The second subset is used to evaluate the method's intrinsic resolution on the electromagnetic energy reconstruction. The result is shown in Figure~\ref{fig:resolution} for different zenith angle bins. The total resolution is below $5\%$, with little variation with zenith angle in the range 57$\degree$-87$\degree$.

\subsection{Performance}
\label{section:energy_perfs}

The performance of the energy reconstruction method is now evaluated on the set of simulations presented in section~\ref{section:simus_realistic}.

\begin{figure}[!ht]
     \centering
    \includegraphics[width=0.49\columnwidth]{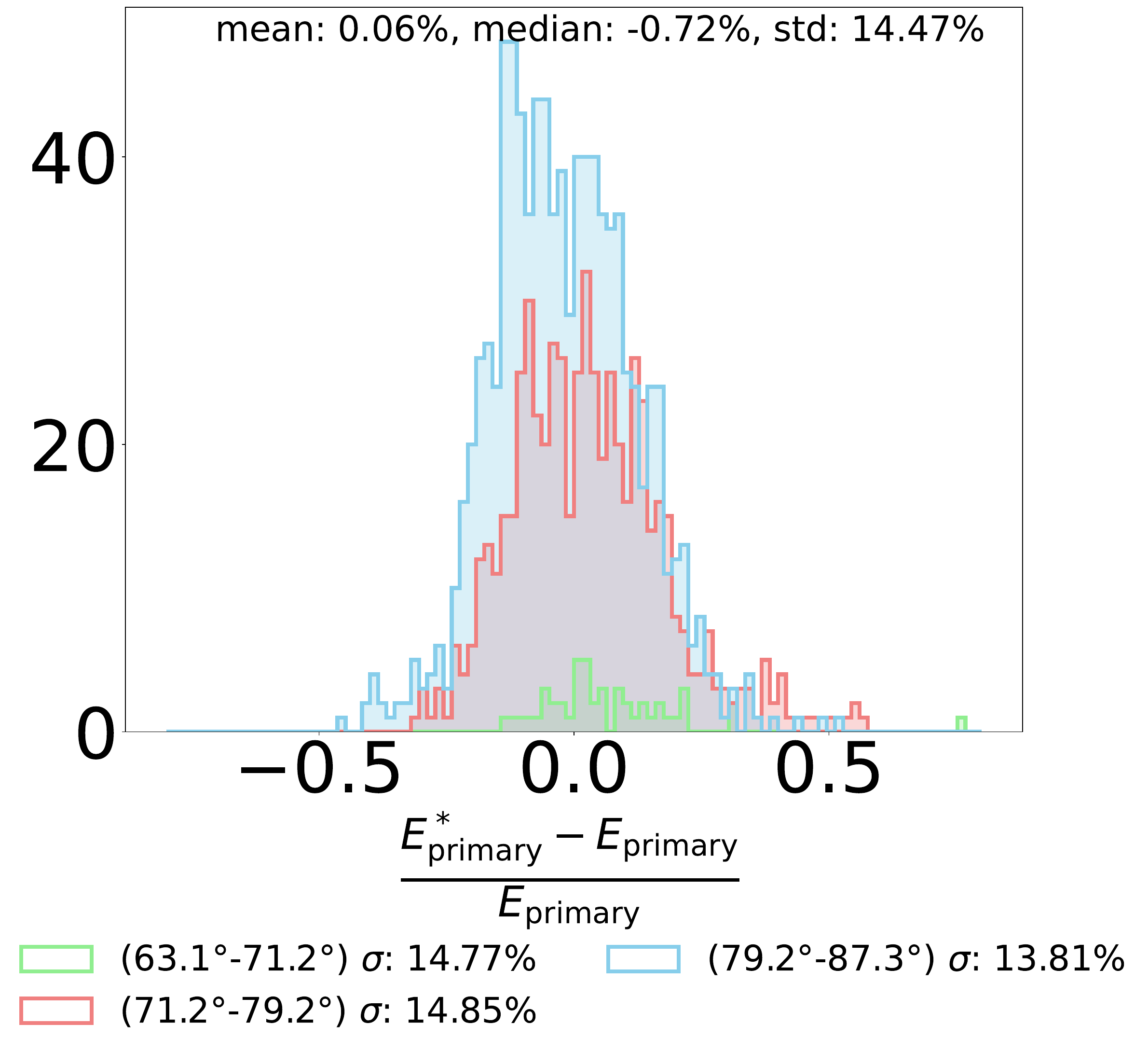}
    \includegraphics[width=0.49\columnwidth]{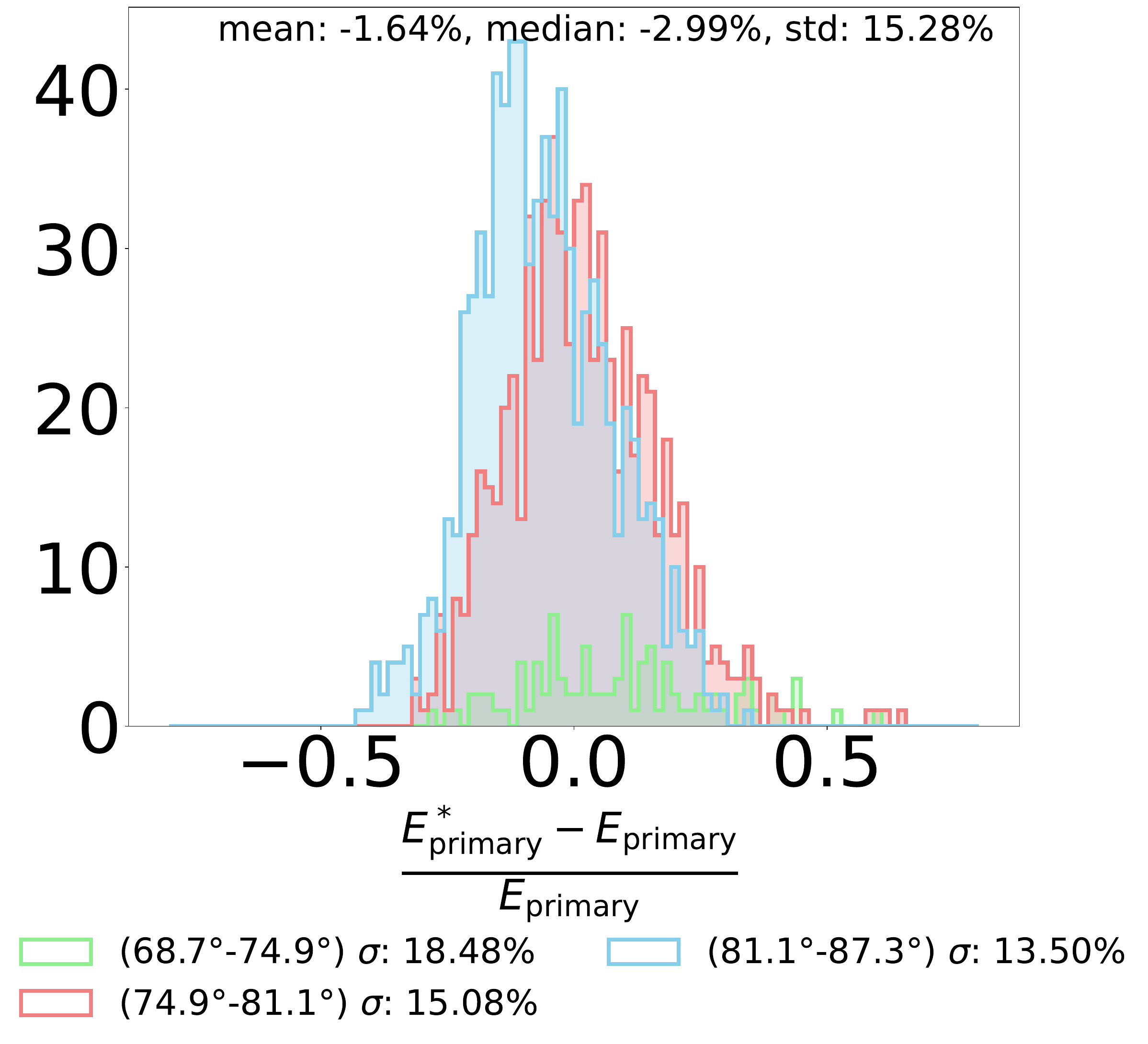}
      \caption{Primary energy resolution for different zenith ranges on the set of realistic simulations  with (\textit{Left}) and without (\textit{Right}) infill.}
    \label{fig:energy_DC2}
\end{figure}

As the electromagnetic energy $E_{\rm em}$ is not accessible in this simulation set, we opt to reconstruct directly the primary energy $E_{\rm primary}$, which is ultimately the parameter of interest. As in the previous section, the simulation set is equally divided into two subsets. The first one is used to compute the density correction factor $f_{\rm primary}(\rho, \rm sin(\alpha))$ and is parameterized as a polynomial regression of degree 3 in $\rho$ and $\rm sin(\alpha)$.
The second set is used to estimate the resolution between the true primary energy and the reconstructed one: 
\begin{eqnarray}
E_{\rm primary}^* = \frac{A}{{\rm sin(\alpha)} f_{\rm primary}(\rho, \rm sin(\alpha))}    
\end{eqnarray}
The results are shown in Figure~\ref{fig:energy_DC2}, for the layouts with and without infills, for which total energy resolutions of 14.5\% and 15.3\% are achieved respectively. The poorer performance in the lower zenith bin for events without the infill is attributed to the low antenna multiplicity (see Figure~\ref{fig:layout}, \textit{Right}). A $\leq$3\% bias on the energy resolution is also observed on the distribution, with a limited variation with zenith angle.
This bias is clearly negligible compared to the achieved energy resolution, and could be related to the composition of the primary particles. Indeed, it should be noted that this study is performed without discrimination of the primary particle type (here iron or proton).
Yet for a given electromagnetic energy, the primary energy varies with the particle's nature. In particular the primary energy fraction carried away by neutrinos and muons (the so-called missing or invisible energy) is larger by a few percent for iron nuclei: 17\% vs 12\% for protons at $10^{17}$\,eV~\cite{Barbosa_2004}.
The quoted 15\% resolution, which combines the method's intrinsic resolution and the fluctuations arising from the nature of the primary particle, may thus be further improved by identifying the nature of the primary particle.

Finally, no significant dependency with energy is observed on the resolution.

\section{Conclusion}
 We have presented in this article a new method to reconstruct  the direction of origin and primary energy of air showers with zenith angles larger than 60$\degree$. Its core element is the {\it Angular Distribution Function}, an analytical description of the amplitude profile expected for radio signals induced by inclined air showers in the 50-200\,MHz frequency range. The ADF is written as a function of the observer's angular position with respect to the shower axis and radio emission point, and is thus applicable even when there is no shower core, i.e. for upward-going showers such as those induced by UHE earth-skimming neutrinos. The ADF is an empirical model, but its various components are motivated by physics argument. This applies in particular to the position of the amplitude maximum, computed with a semi-analytical model of the Cherenkov emission presented in this article. There are consequently only 4 free parameters (direction ($\theta, \phi$), amplitude and width of the Cherenkov peak) for the adjustment of the ADF to the actual amplitude distribution. The radio emission point used in the ADF is reconstructed from the arrival times of the EAS following the work presented in~\cite{decoene_2023} and is correctly suited for inclined air showers.

We have tested the ADF reconstruction method on simulated data in the zenith range 60-87$\degree$, assuming realistic experimental conditions: stationary noise on the radio signals corresponding to sky emission in the 50-200\,MHz frequency range, 5\,ns resolution on timing and 7.5\% fluctuations on signal amplitude, corresponding to what can be realistically achieved through calibration. We have shown that the shower direction of origin can be reconstructed on this simulation dataset with a mean resolution better than 4 arc-minutes ($0.07\degree$) for events with at least 6 antennas triggered. An intrinsic resolution better than $5\%$ is obtained on the electromagnetic energy, while a resolution better than 15\% is achieved for the primary energy. This method is thus well suited to achieve the goals of experiments composed of sparse radio arrays targeting UHE cosmic particles, such as GRAND or AugerPrime  radio upgrade. Full reconstruction performances estimate would require that the response of the specific detectors is taken into account to compute the electric field signals from the experiments data, a procedure that could be performed by these collaborations.

The present work shall also be completed in a near future by studying the method performances for non-core contained and upward-going EAS.

\section*{Acknowledgments}
We thank GRAND collaborators (and in particular M. Bustamante, K. Kotera, T. Bister) for useful discussions and comments. This work is  supported by the ANR (ANR-21-CE31-0025) and the DFG (Projektnummer
490843803). Simulations were performed using the computing resources at the IN2P3
Computing Centre (Lyon, France), a partnership between CNRS/IN2P3 and CEA/DSM/Irfu.

\appendix

\section{Performance of the arrival direction reconstruction procedure}
\label{appendix:direction}
The first two steps of the ADF reconstruction procedure involve adjusting the radio wavefront for both plane and spherical waves. It is worthwhile to evaluate the performance of these quick and straightforward methods by comparing certain observables to their true values. We conduct this evaluation using the simulation set presented in section \ref{section:simus_realistic}. The statistics indicate that 96\% of the events resulted in a successful plane wave fit.

\begin{figure}[!ht]
     \centering
    \includegraphics[width=0.49\columnwidth]{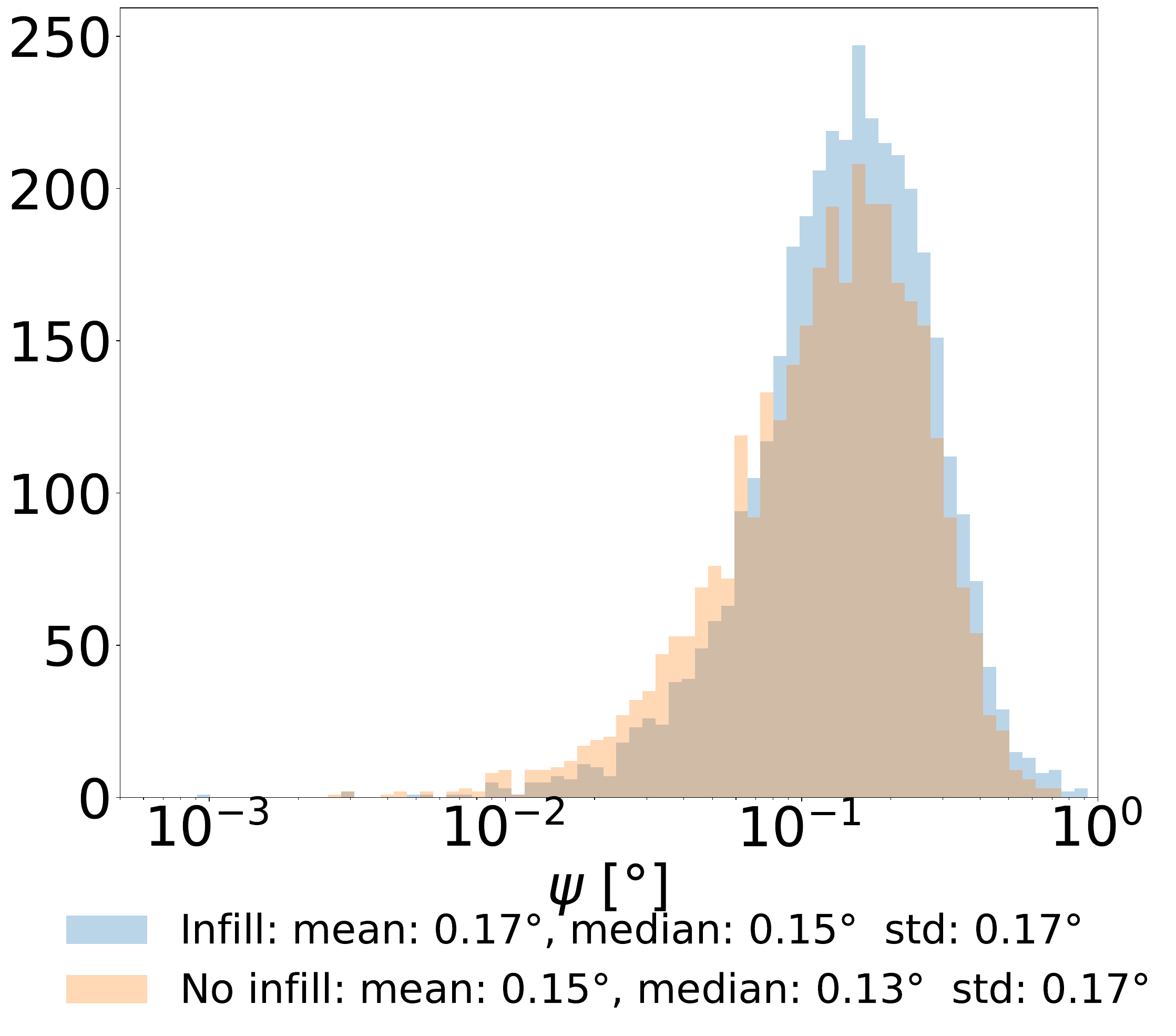}
    \includegraphics[width=0.49\columnwidth]{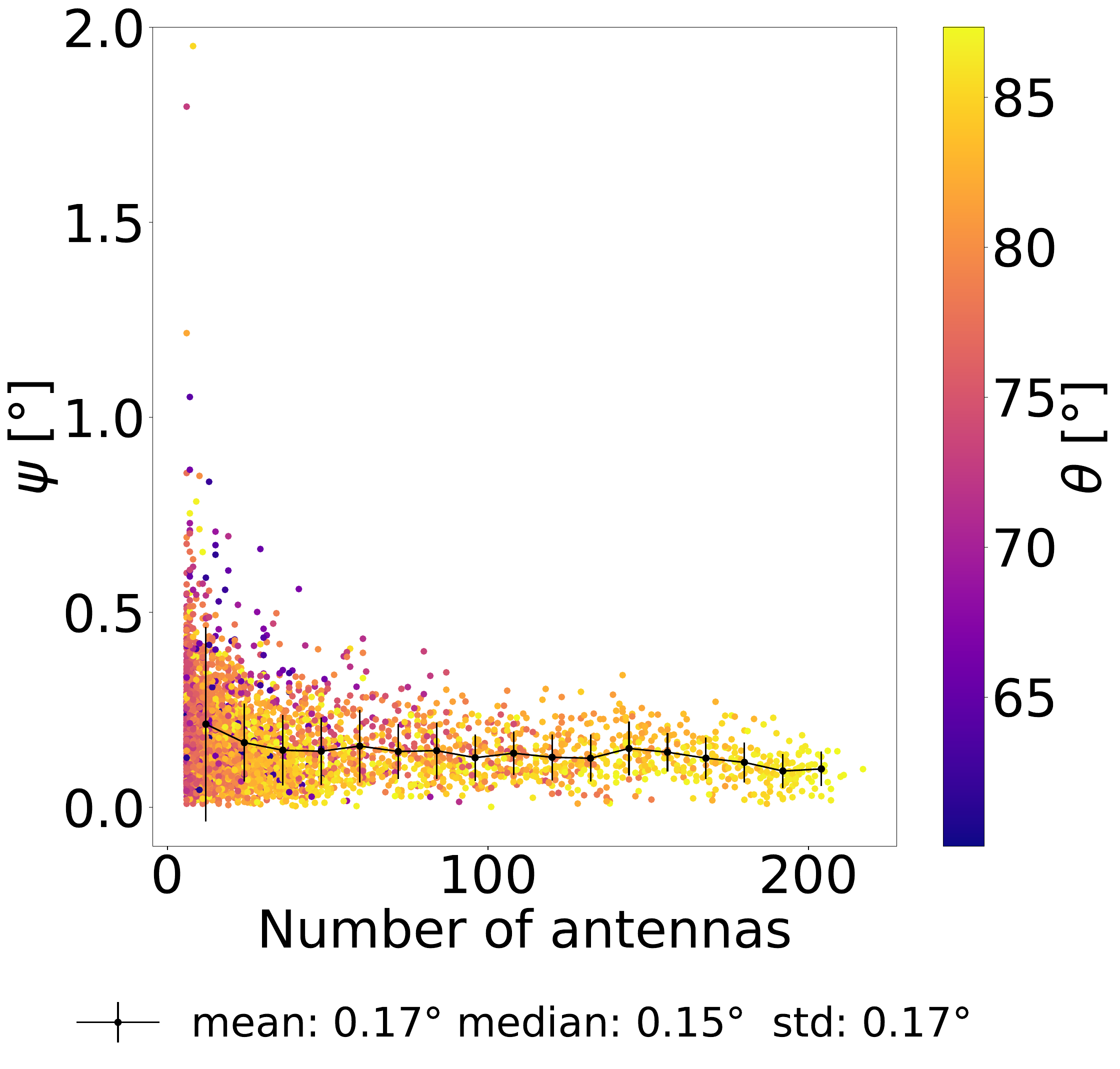}

      \caption{(\textit{Left}) Distribution of the angular distance $\psi$ between the true shower direction and the reconstructed obtained for the plane wave reconstruction. Results for the hexagonal layout with (without) infill are shown in blue (orange).  (\textit{Right}) Angular distance $\psi$ as a function of number of antennas and inclination $\theta$ for the plane wave reconstruction for the hexagonal layout with infill.}
    \label{fig:PWF}
\end{figure}

\begin{figure}[!ht]
     \centering
    \includegraphics[width=0.49\columnwidth]{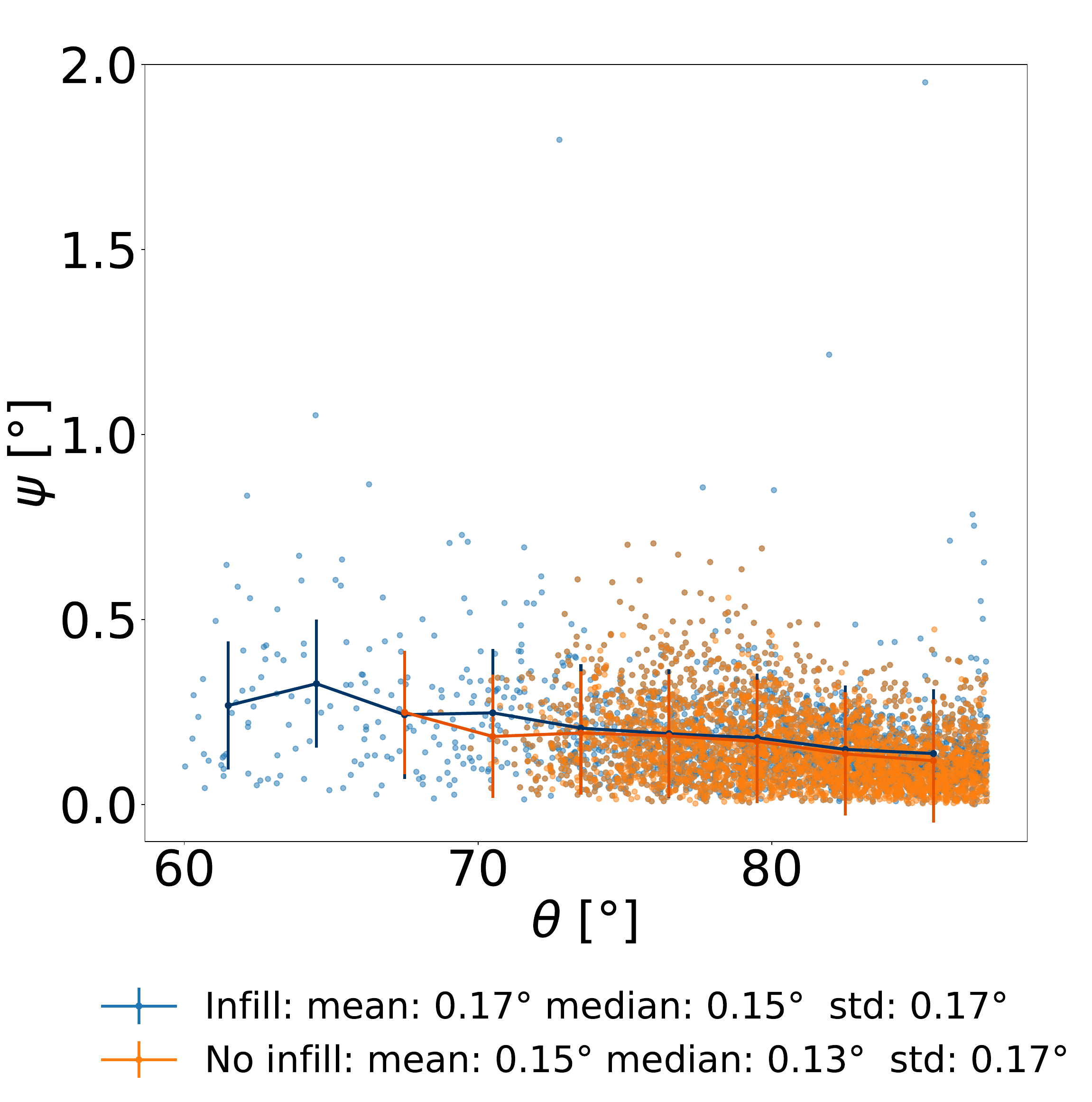}
    \includegraphics[width=0.49\columnwidth]{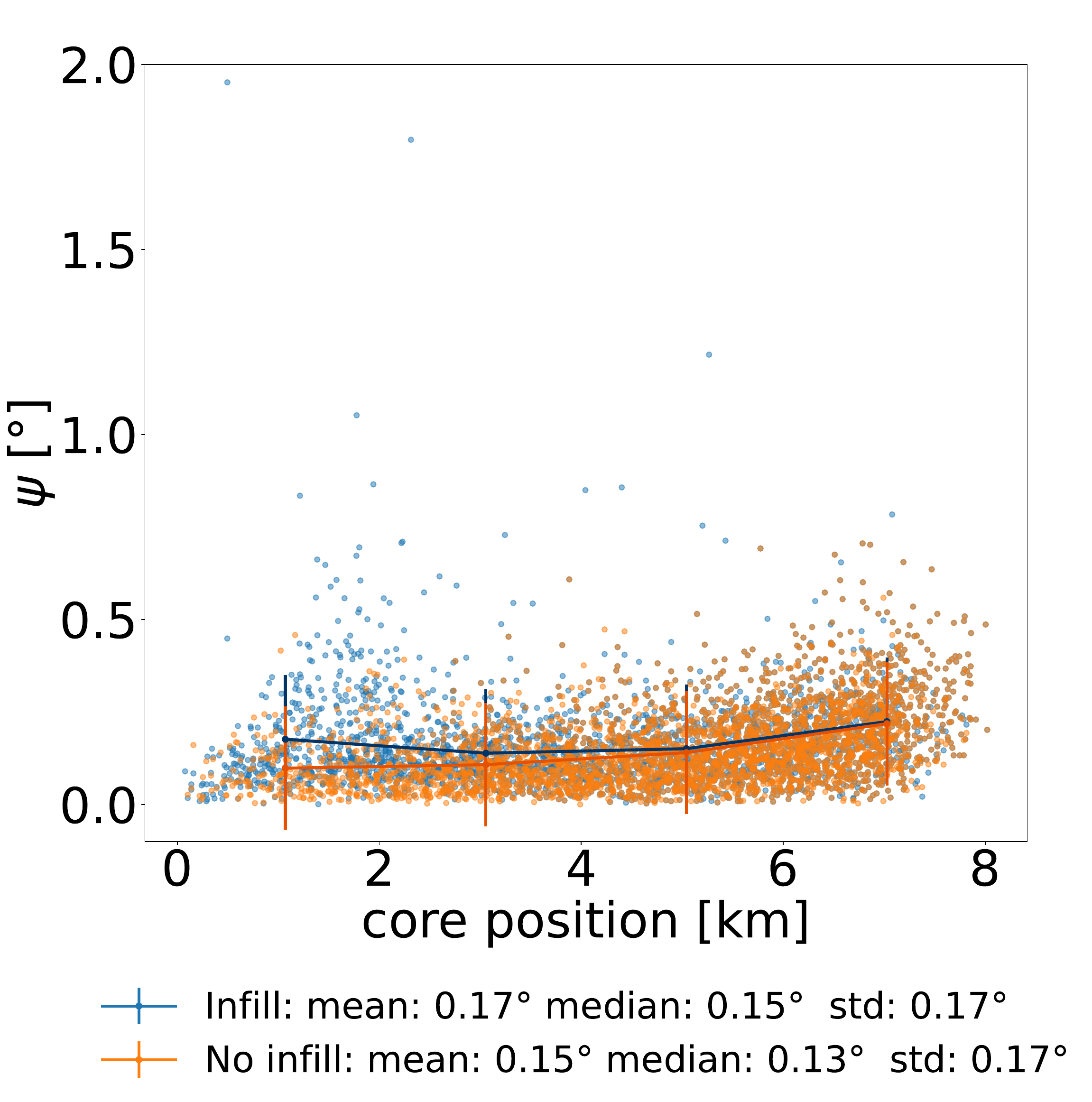}
      \caption{(\textit{Left}) Angular distance $\psi$ between the true and reconstructed shower direction for the plane wave fit.  (\textit{Right}) Angular distance $\psi$ as a function of core position for the plane wave reconstruction. Results for the hexagonal layout with (without) infill are shown in blue (orange). The main array spans a hexagon with side length 8.1 km, while the infill array covers a smaller hexagon with side length 1.625 km. In both plots only events with $N_{\rm ants} \geq 6$} are shown.
    \label{fig:PWF_stats}
\end{figure}

We first display in Figures~\ref{fig:PWF} and \ref{fig:PWF_stats} the angular resolution obtained with the plane wave reconstruction as a function of a number of triggered antennas, inclination, and core positions. The resolution obtained is better than $0.2\degree$, a remarkable result that opens exciting perspectives for prompt (online) reconstruction. Yet it can be observed that the resolution degrades faster than the ADF for smaller zenith angles (despite low statistics) or number of antennas, and peripheral core positions. 
Simulations with a shower core outside the detector area are expected to negatively impact the results of the plane wave construction.

\begin{figure}[!ht]
     \centering
    \includegraphics[width=0.49\columnwidth]{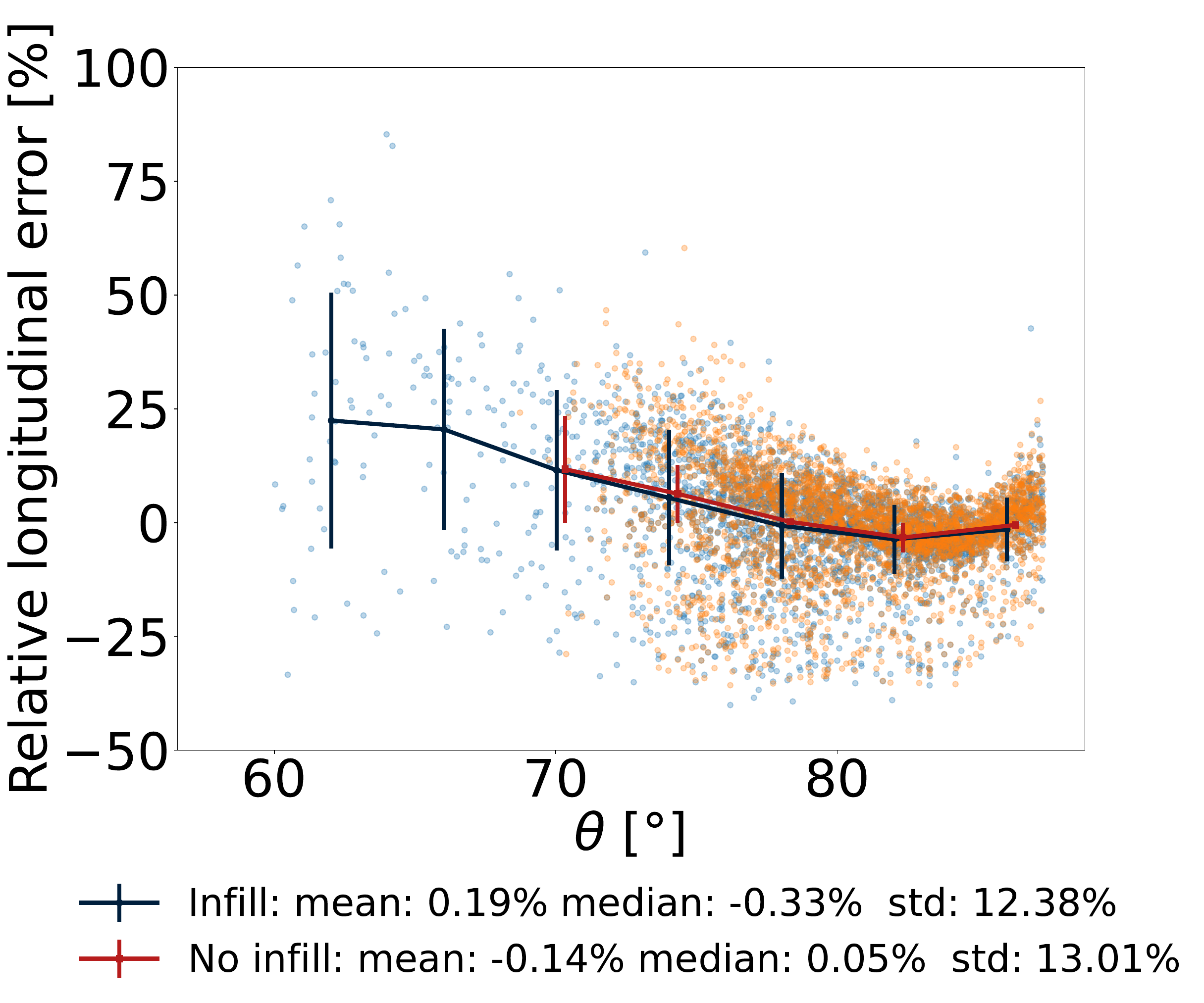}
    \includegraphics[width=0.49\columnwidth]{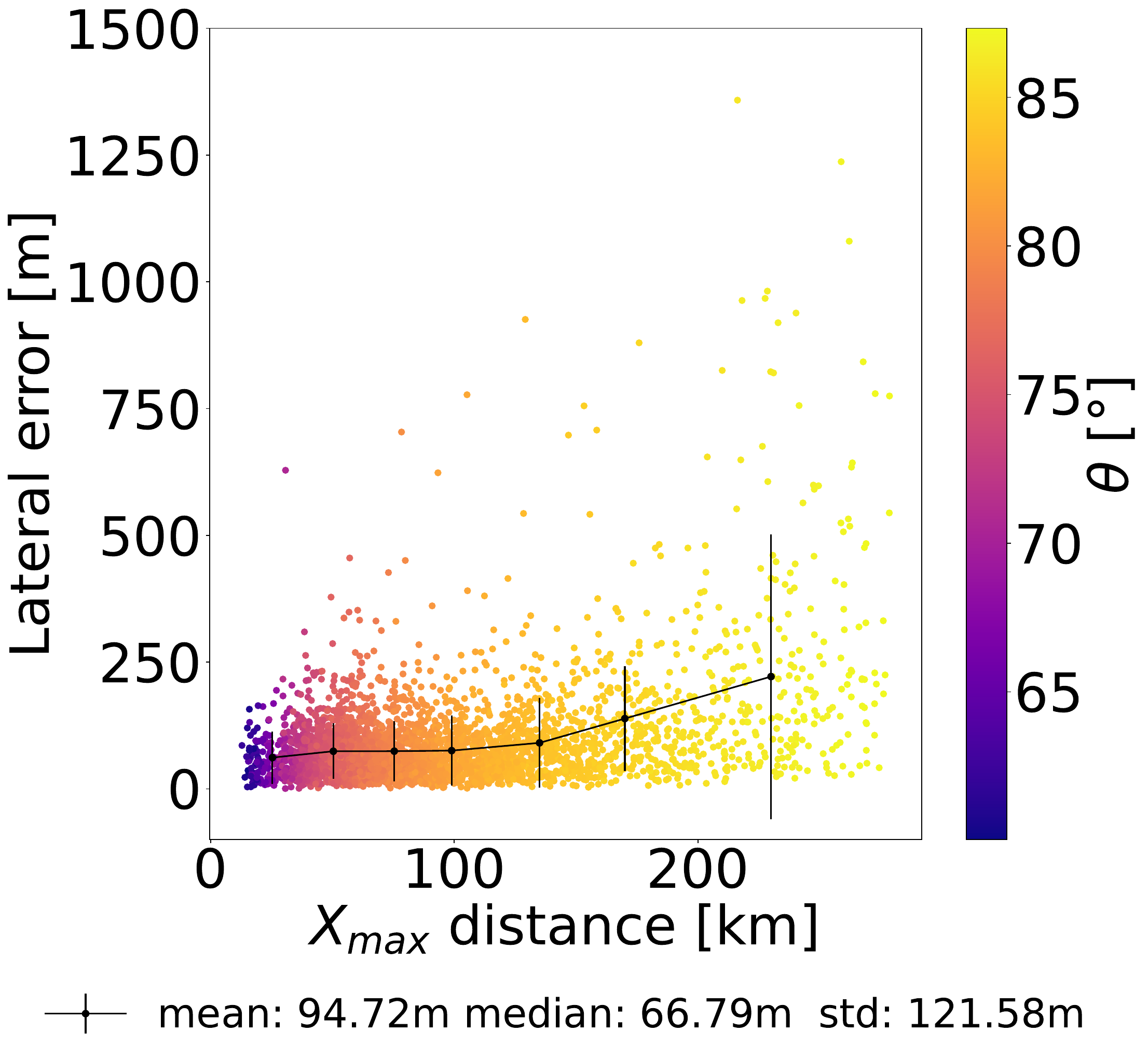}
      \caption{(\textit{Left}) Longitudinal error between the true $X_{\rm max}$ position and the reconstructed $X_{\rm e}$ position. Results for the hexagonal layout with (without) infill are shown in blue (orange). (\textit{Right}) Lateral error, defined as the distance of the reconstructed $X_{\rm e}$ position to the shower axis with infill.}
    \label{fig:SWF_stats}
\end{figure}

Figure~\ref{fig:SWF_stats} shows the relative longitudinal error (left) and lateral error (right) as a function of inclination for the spherical reconstruction. The relative longitudinal error is computed as the difference between the $X_{\rm max}$ position and the emission point-source $X_{\rm e}$ along the shower axis, normalized by the distance between the $X_{\rm max}$ location and the shower core. Interestingly, this error decreases with inclination, which clearly shows that the point source assumption behind the spherical wavefront reconstruction holds better at high inclination. Of course, this is expected since the distance between the array and the $X_{\rm max}$ location increases drastically as the trajectories becomes more horizontal ($\propto 1/\cos{\qty(\theta)}$). Consequently, the emission source is located further away and the true wavefront curvature becomes more and more spherical with propagation. Furthermore, the relative error also reduces because the total longitudinal distance increases hence reducing the ratio.
The lateral error is the distance in the orthogonal direction to the true shower axis between the $X_{\rm max}$ position and the reconstructed $X_{\rm e}$ position. Its relative mean remains stable with inclination, ranging from $0.1\%$ at $\theta = 65\degree$ to $0.08\%$ at $\theta = 85\degree$. This error is relatively small in comparison to the longitudinal distance, which can be explained by the powerful lever arm provided by the curvature of the wavefront. While the longitudinal error is sensitive to the absolute time of arrival of the wavefront model, the lateral error is sensitive to the relative time of arrival between antennas. This relative time is directly related to the wavefront curvature, which is accurately described by the spherical wavefront model (taking into account our typical time resolutions).

\section{Electromagnetic energy computation from ZHAireS}
As only electrons and positrons are associated with electromagnetic emission of EAS detected by radio antennas, the amplitude of the radio emission is directly linked to the so-called shower electromagnetic energy $E_{\rm em}$, defined as the total energy released by the electromagnetic component of the shower: electrons, positrons, and gammas.

We calculate the electromagnetic energy from the ZHAireS simulations by summing the longitudinal energy deposited by electrons and positrons through ionization, denoted as $(E^{\rm ionization}_i)$, along with the longitudinal energy of the discarded electrons, positrons, and gammas that fall below the energy threshold $(E^{\rm cut}_i)$. It is reasonable to assert that all low-energy particles deposit their energy. Additionally, we include the energy deposited in the ground plane, $(E^{\rm ground})$, by electrons, positrons, and gammas. However, since we focus on inclined air showers with an angle $(\theta > 60\degree)$, the showers can fully evolve before the particles reach the ground. Consequently, the energy deposited at the ground is negligible for these shower geometries and will not impact the total electromagnetic energy calculation.

Finally, the true Monte Carlo electromagnetic energy derived from the ZHAireS simulations can be written as:
\begin{eqnarray}
\label{eq:em_energy}
E_{\rm em} = \sum^{N}_{i=0} E_{i}^{\rm ionization} (e^{+}e^{-}) +  E_{i}^{\rm cut} (e^{+} e^{-} \gamma) + E^{\rm ground} (e^{+} e^{-} \gamma) \ .
\end{eqnarray}
\label{sec:em_energy}
where $i$ refers to the longitudinal profile bins.

\section{Toy model description of the Cherenkov angle}
\label{sec:annex:TMCherenkov}
In this appendix, we first outline the phenomenological framework of the toy model used to calculate the Cherenkov angle. This calculation focuses explicitly on very inclined air showers and considers the shower's geometry and the observer's location. Next, we provide a detailed analytical derivation to implement this computation. This model should be viewed as a preliminary and superficial exploration of the physics of radio Cherenkov effects in very inclined air showers. It is motivated by the need for predictive values of the Cherenkov angle to serve as input for a reconstruction algorithm rather than as a comprehensive description of the underlying physical processes. A more accurate model and derivation will be presented in a future work.

\paragraph{Phenomenological description}
In our toy model, we choose two emission points along the EAS track, placed around the maximum development of the shower, separated by a distance $\Delta \sim1-2$\,km (see a sketch in Figure~\ref{fig:sketch_cherenkov_model}). Note that the value of $\Delta$ is chosen according to typical longitudinal shower developments for inclined trajectories~\cite{guelfand_2024}. We then compute the time delay measured by an observer at the ground in the reception of the signals emitted from these 2 points and determine how this time delay varies with the observer's position. 
The exact value (in a reasonable range) of $\Delta$ only induces a second-order effect and does not change the results significantly, since we are only interested in the comparison between the two emission points. Finally, the effective refraction index of each observer's line of sight is taken into account.

\begin{figure}[!ht]
   \centering   
   \includegraphics[width=0.49\linewidth]{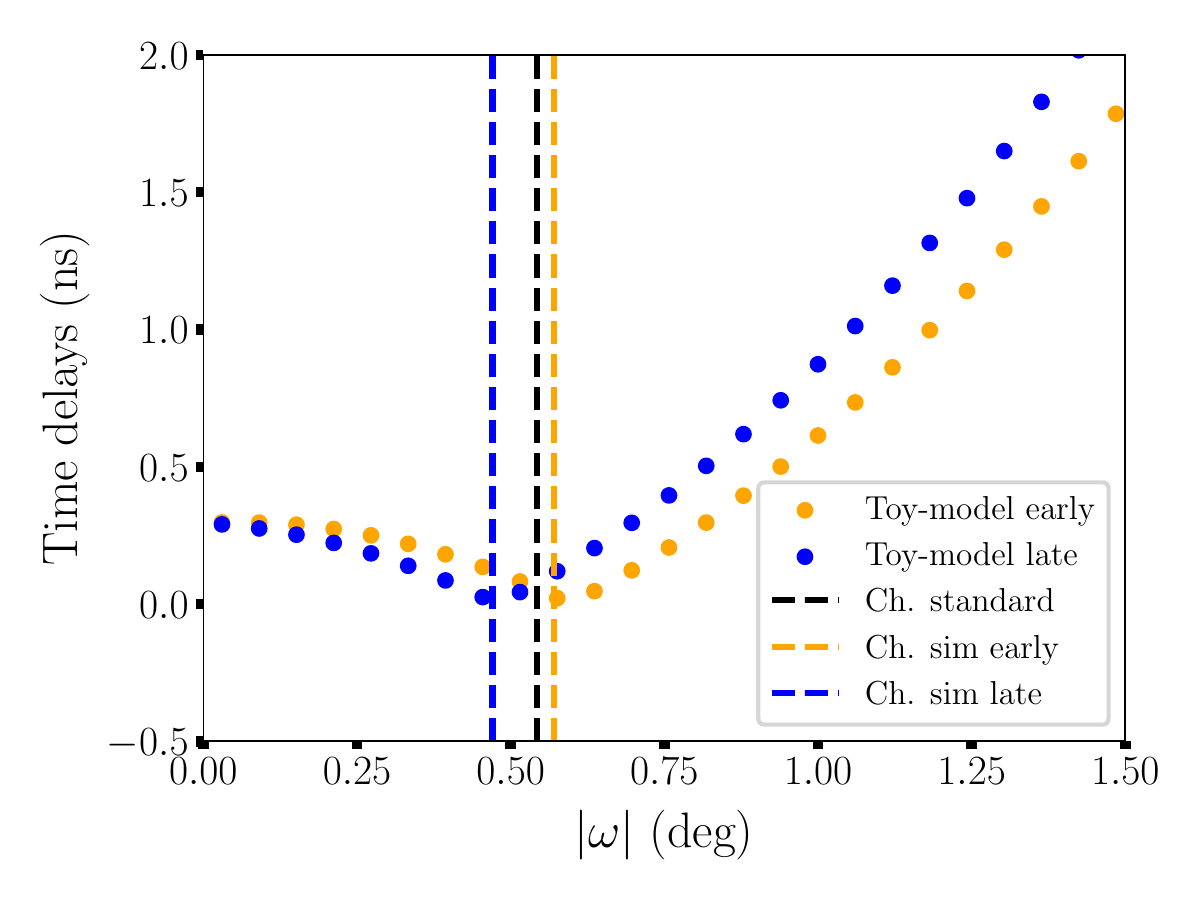} 
    \includegraphics[width=0.49\linewidth]{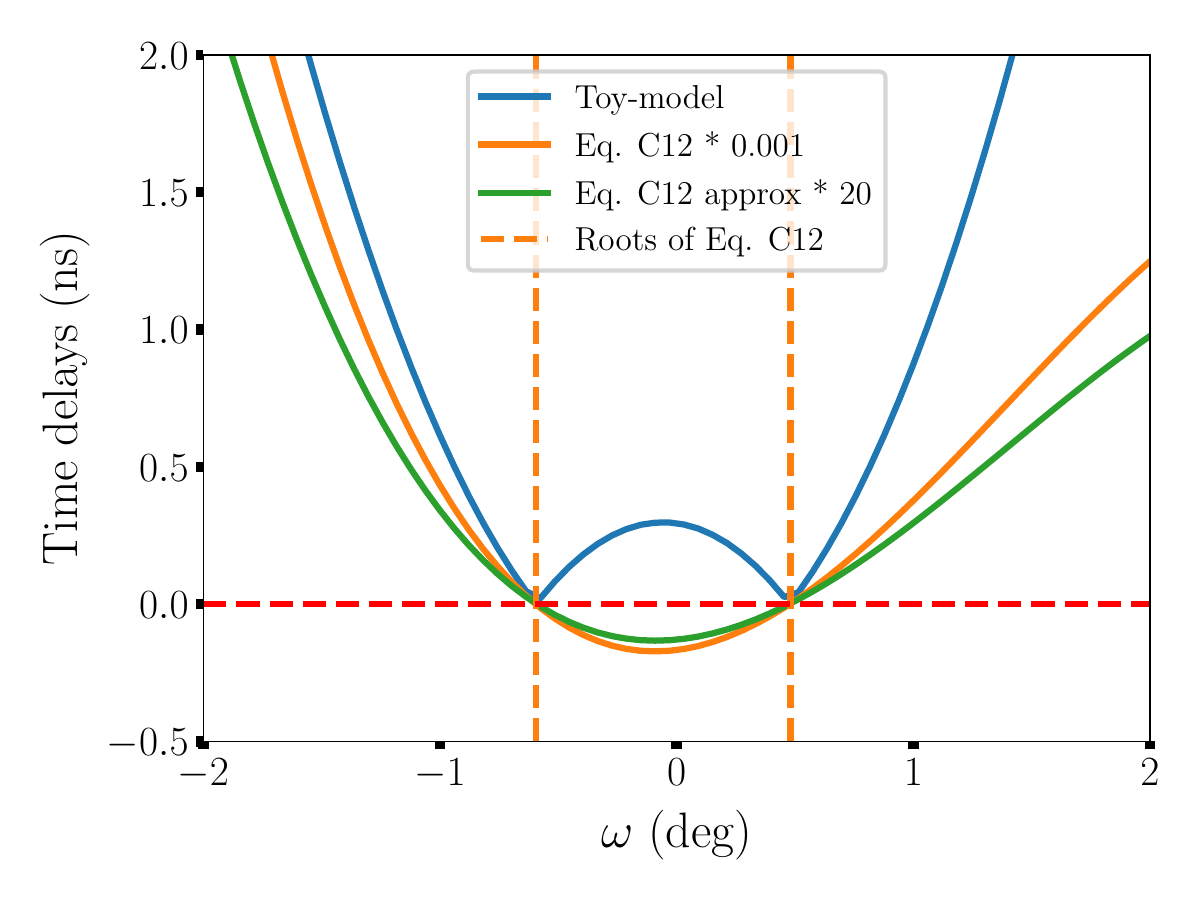} 
   \caption{{\it Left: }Time delays as a function of observer position, computed from the two emission points of the toy model for an EAS with direction $(\phi, \theta) = (0\degree, 85.8\degree)$, and energy $E=4.92$\,EeV. We can clearly see two Cherenkov angles for early (orange) and late (blue) antennas. The standard Cherenkov angles computed from the refractive index at the altitude of the two emission points fails to reproduce the exact Cherenkov values as well as the asymmetry between early and late observers. 
   {\it Right:} In blue, the toy-model time delays, in orange, Equation~\ref{eq:equation} and in green Equation~\ref{eq:4}. From the figure we can infer a quartic polynomial for Equation~\ref{eq:equation} and Equation~\ref{eq:4} the roots of which exactly match the minima of the time delays. Vertical lines are the solutions of Equation~\ref{eq:equation} found by a basic dichotomy search. Figure taken from~\cite{decoene_thesis2020}.}
   \label{fig:time_delays_equations}
\end{figure}

As will be shown below, it is possible to compute numerically the time delays between these two points, using the same atmosphere model as ZHAireS for the computation of the refractive index and signal propagation time. The Cherenkov angle can then be associated with the angle value for which the time delays equals zero (see~\cite{schroder_2016}).
Figure~\ref{fig:time_delays_equations} (left panel) displays the time delay values for various observer positions, for an EAS with direction $(\phi, \theta) = (180\degree, 85.8\degree)$, and energy $E=4.92$\,EeV. The values of the Cherenkov angle (for which time delays are null) clearly differ for early and late antennas.

This simple toy model enables us to replicate the asymmetry effect observed between early and late antennas. The next step is to develop an analytical model that will allow us to compute this effect for various shower configurations quantitatively. This will help us extract the Cherenkov angle values for any shower geometry.

\paragraph{Analytical model}
Consider the time delay between the optical path from a given point $E$ (taken above the $X_{\rm max}$ position) to an observer position $\omega$ and the one from $X_{\rm max}$ to that same observer position $\omega$. Note that $\omega<0$ corresponds to early antennas and $\omega>0$ to late antennas. The two lines of sight starting from the observer and reaching $E$ or the $X_{\rm max}$ position are denoted $l_1$ and $l_0$ respectively.
To each of these lines of sight $l_{1}$ and $l_{0}$, corresponds
a given effective index of refraction $n_1$ and $n_0$, defined as the mean value of the refractive index along the line of sight. The time delay can then be written,
\begin{align} \label{eq:1}
c \delta t = n_{0} l_{0} - n_{1} l_{1} + \Delta \ ,
\end{align}
where $l_{0}$ is given by the solution of
\begin{align}
l_0^2 &= L_{X}^2 + d^2 - 2  L_{X} d \cos{\qty(\alpha)}\ ,
\\ d^2 &=   L_{X}^2 + l_0^2 - 2  L_{X} l_0 \cos{\qty(\omega)}\ ,
\end{align}
where $d$ is the distance between the core and the observer position, $L_X$ the distance to $X_{\rm max}$ from core, and $\alpha$ the angle between the shower axis and the ground ($\geq\pi/2$), leading to
\begin{align}  \label{eq:2}
l_{0} = \frac{ L_{X} \sin{\qty(\alpha)}}{\sin{\qty(\alpha + \omega)}} \ .
\end{align}
After solving a second-order polynomial in $l_0$ and a few simplifications in terms of sine and cosine (or via the Al-Kashi theorem directly), we obtain
\begin{align}   \label{eq:3}
l_{1} = \sqrt{l_{0}^2 + \Delta^2 + 2 \Delta l_0 \cos(\omega)} \ .
\end{align}
The Cherenkov angle $\omega_{\rm C}$ is the solution to $c\delta t \approx 0$ or equivalently $c^2\dv{ (\delta t)^2}{\omega} = 0$.
In the latter case, the derivative of Equation~\ref{eq:1} squared is given by
\begin{align}
c^2\dv{ (\delta t)^2}{\omega} = 2 \qty(n_{0} l_{0} - n_{1} l_{1} + \Delta) \qty(n_0 \dv{l_0}{\omega} - n_1\dv{l_{1}}{\omega}) \ ,
\end{align}
which only cancels out for $n_{0} l_{0} - n_{1} l_{1} + \Delta = 0 = c\delta t$, since the time delays are strictly growing functions of $\omega$. We can perform a limited expansion of Equation~\ref{eq:3} in terms of $\Delta / l_{0} \ll 1$ inside Equation~\ref{eq:1}, which gives
\begin{align}
l_0 \qty(n_0 - n_1) + \Delta \qty(1 - n_1 \cos{\omega}) = 0 \ 
\end{align}
Then replacing $l_0$ by its expression in Equation~\ref{eq:2} yields
\begin{align} \label{eq:4}
L_{X} \sin{\qty(\alpha)} \qty(n_0 - n_1) + \Delta \qty(1 - n_1 \cos{\omega}) \sin{\qty(\alpha - \omega)} = 0 \ .
\end{align}
The equation is satisfied for $\omega = \omega_{\rm c}$, and cannot be solved analytically.

Since a numerical solution is needed for the approximate expression, let us look for an exact computation. It can be achieved by looking at the square of Equation~\ref{eq:1}
\begin{align}
(n_0 l_0 + \Delta )^2 = (n_1 \, l_1)^2 \ ,
\end{align}
leading to 
\begin{align}
l_0^2 \qty(n_0^2 - n_1^2) + \Delta^2 \qty(1 - n_1^2) + 2 l_0 \Delta \qty(n_0 - n_1^2 \cos{\qty(\omega)}) = 0 \ .
\end{align}
Replacing $l_0$ by its expression in Equation~\ref{eq:2} yields
\begin{align} \label{eq:equation}
L_{X}^2 \sin{\qty(\alpha)}^2 \qty(n_0^2 - n_1^2) &+ \Delta^2 \qty(1 - n_1^2) \sin{\qty(\alpha + \omega)}^2 \nonumber
\\ &+ 2 L_X \Delta \qty(n_0 - n_1^2 \cos{\qty(\omega)}) \sin{\qty(\alpha)}\sin{\qty(\alpha + \omega)} = 0 \ .
\end{align}
This equation hides a quartic polynomial, which can be developed as a function of $t = \tan{\qty(\omega/2)}$, under the form,
\begin{align}
- &t^4 \, \sin{\qty(\alpha)} \qty[2 l_0 n_0 \Delta + n_1^2] \nonumber
\\&- t^3 \, 2 \cos{\qty(\alpha)} \qty[2 l_0 n_0 \Delta + n_1] \nonumber
\\ &+ t^2 \, \sin{\qty(\alpha)} \qty[n_1^2 \qty(2 + \Delta^2 - L^2) + n_0^2 L_{X}^2 - \Delta] \nonumber
\\ &+ t \, 2\cos{\qty(\alpha)} \qty[n_1^2 \qty(1 + \Delta^2) - \Delta^2 - 2 l_0 n_0 \Delta] \nonumber
\\ &+ \sin{\qty(\alpha)} \qty[2 l_0 n_0 \Delta + \Delta^2 - n_1^2 \qty(\Delta^2 + 1 + L_{X}^2) + L_{X} n_0^2] \nonumber
\\ &= 0 \ ,
\end{align}
with no obvious solutions (see Figure~\ref{fig:time_delays_equations}, right panel). At this stage, a numerical solution is needed, following, for example, a basic dichotomy search. This treatment allows us to determine a numerical value for the Cherenkov position from the shower geometry only. Interestingly, in the case where the refractive index has a constant profile with altitude, it converges to the standard computation $\omega_c=\arccos{\qty(1/n)}$. For example,
at $\theta=80\degree$ we find with the toy-model $\omega_c^{\rm n=cste}=0.64\degree$, (while in the case of a realistic profile $\omega_c^{\rm n=f(h)}=-0.62/0.67\degree$ for the early and late angles respectively) and the standard computation gives $\omega_c^{\rm standard}=0.64\degree$. Another example at $\theta=85\degree$ gives $\omega_c^{\rm n=cste}=0.56\degree$ ($\omega_c^{\rm n=f(h)}=-0.49/0.62\degree$) and $\omega_c^{\rm standard}=0.56\degree$. We clearly see how both the toy-model and the realistic refractive index profile are required to reproduce the observed asymmetry in the simulations.
Note that an independent treatment of this asymmetry was published after this work was carried out~\cite{Schluter_2020}.
However, it was performed on events less inclined than in this study.

This model allows for a treatment of the Cherenkov asymmetry in the modeling of the amplitude distribution. For a practical implementation we can generalize to any observer location by computing the $\eta$ angle (defined in the shower plane, see Fig.~\ref{fig:sketch_adf_coordinates_systems}), and proceed to the following substitutions
\begin{align}\label{eq:substitutions}
    \omega &= \omega {\rm\, sign}\qty({\eta}) \nonumber
    \\ \alpha &= \arccos{\qty(\cos{\qty(\eta)}\sin{\qty(\theta)})} \ .
\end{align}
By replacing Eqs.~\ref{eq:substitutions} int Eq.~\ref{eq:equation}, we can compute the expected Cherenkov angle from any observer location.

\section{Minimization procedure}
\label{annex:mini}
For the reconstruction of plane waves, spherical waves, and ADF, we rigorously explored a variety of methodological approaches. These included optimization using the \texttt{scipy.optimize} library with both numerical and analytical Jacobians, as well as packages built upon \texttt{scipy}, such as \texttt{lmfit}\footnote{\url{https://lmfit.github.io/lmfit-py/}}, which is particularly well-suited for handling bounded parameters. Additionally, we evaluated Markov Chain Monte Carlo methods, which provided robust results but were prohibitively slow for our use cases. Ultimately, we adopted a fully analytical approach for plane wave reconstruction, while for the Spherical Wavefront reconstruction, we used the differential evolution method from the \texttt{scipy.optimize} library. For the ADF reconstruction, we utilized the \texttt{MINUIT} algorithm through its Python wrapper, \texttt{iminuit}\footnote{\url{https://scikit-hep.org/iminuit/}}. This choice was motivated by the algorithm's ability to robustly account for the covariance of the model parameters. The four free parameters of the ADF fit are constrained within specified boundaries to ensure better convergence. Based on the plane wave reconstruction, we restrict the values of $\theta$ and $\phi$ to the intervals $\theta \in [\theta_{\rm plan} - 2^\circ, \theta_{\rm plan} + 2^\circ]$ and $\phi \in [\phi_{\rm plan} - 1^\circ, \phi_{\rm plan} + 1^\circ]$ \cite{decoene_thesis2020}. The scaling factor $A$ is constrained to the range $[10^6, 10^{10}]$, while the Lorentzian width $\delta \omega$ is restricted to the interval $[1.25, 3]$.

\bibliographystyle{elsarticle-num-names}
\bibliography{biblio.bib}

\end{document}